\documentclass[12pt]{article}
\usepackage{amsmath, amsfonts}
\usepackage{graphicx,psfrag,epsf}
\usepackage{enumerate}
\usepackage{natbib}
\usepackage{float}
\usepackage{url} 
\usepackage{enumitem}
\setlist{nolistsep}

\newcommand{\blind}{0}

\begin{document}

\def\spacingset#1{\renewcommand{\baselinestretch}%
{#1}\small\normalsize} \spacingset{1}

\if0\blind
{
  \title{\bf Trend and Variance Adaptive \\ Bayesian Changepoint Analysis \\ and Local Outlier Scoring}
  \author{Haoxuan Wu\thanks{Corresponding author: hw399@cornell.edu.}, \hspace{.1cm} Toryn L.J. Schafer and David S.\ Matteson \thanks{Financial support is gratefully acknowledged from a Xerox PARC Faculty Research Award, National Science Foundation Awards 1455172, 1934985, 1940124, 1940276 and 2114143, USAID, and Cornell University Atkinson Center for a Sustainable Future.}\hspace{.2cm} \thanks{The authors would like to thank Dr.\ Michael Jauch, Post-Doctoral Associate at Cornell Center for Applied Mathematics, for his detailed feedback on early drafts of this manuscript.}\hspace{.2cm} \thanks{Declaration of Interest: None} \\
    Department of Statistics and Data Science, Cornell University}
  \maketitle
} \fi

\if1\blind
{
  \bigskip
  \bigskip
  \bigskip
  \begin{center}
    {\LARGE\bf Trend and Variance Adaptive Bayesian Changepoint Analysis \& Local Outlier Scoring}
\end{center}
  \medskip
} \fi

\bigskip
\begin{abstract}
We adaptively estimate both changepoints and local outlier processes in a Bayesian dynamic linear model with global-local shrinkage priors in a novel model we call Adaptive Bayesian Changepoints with  Outliers (ABCO). We utilize a state-space approach to identify a dynamic signal in the presence of outliers and measurement error with stochastic volatility. We find that global state equation parameters are inadequate for most real applications and we include local parameters to track noise at each time-step. This setup provides a flexible framework to detect unspecified changepoints in complex series, such as those with large interruptions in local trends, with robustness to outliers and heteroskedastic noise. Finally, we compare our algorithm against several alternatives to demonstrate its efficacy in diverse simulation scenarios and two empirical examples on the U.S. economy.
\end{abstract}

\noindent%
\vfill

\newpage
\spacingset{1.8} 

\section{Introduction}
Changepoint analysis involves detecting instantaneous changes in the distribution of a time series. This field has a wide variety of applications: in genetics, it is used to identify changes within DNA sequences \citep{Braun_2000}; in environmental science, it is applied to quantify climate change \citep{Solow_1987}; in finance, it helps gain insights into historical data and improves future forecasting \citep{Chen_1997}; in economics, changepoint analysis of key economic indicators illustrates periods of changing conditions and reactions to policy \citep{rahbek2002autoregressive, sims2006were, giordani2008efficient, hauzenberger2022fast}. As these examples illustrate, modern time series is characterized by increased complexity and decreased homogeneity \citep{Rehman_2016}. We aim to understand how the underlying distribution of a time series changes, to distinguish unspecified local trends from major changes while accounting for stochastic volatility and outliers. The flexible and adaptive framework we propose can extend the application of changepoint analysis to even more domains of research with complex time series.  

The proposed approaches to identify changepoints are numerous. One segment of changepoint literature uses various types of statistical criterion such as cumulative sum \citep{Cho_2015}, energy distance \citep{Matteson_2013} or Kullback-divergence \citep{Liu_2013} to segment a series into contiguous clusters. While these approaches have shown effectiveness in estimating the number and locations of changepoints, they have limited flexibility and do not typically provide uncertainty quantification. On the other hand, one can also find changepoint detection methods based on recurrent neural networks \citep{Ebrahimzadeh_2019} or Gaussian processes \citep{Saatci_2010}. Such approaches can be extremely flexible, but may lack robustness or interpretability due to their `black-box' nature. They may also require large training series, expert structural specification, and extensive hyper-parameter selection. 

From the Bayesian perspective, changepoint locations can be incorporated through mixture models, discrete priors, or considered a posterior summary \citep{mcculloch1994bayesian, Adams_2007, wu2024drift}. \citet{mcculloch1994bayesian} introduced a Gibbs sampling algorithm for a random level-shift autoregressive model characterized by a latent binary indicator. Hidden Markov models track changes of latent states using Markov transition matrices \citep{rabiner1989tutorial, luong2012hidden}. Often Dirichlet processes are utilized as priors for the transition matrices \citep{maheu2016infinite, ko2015dirichlet} or structural parameters \citep{peluso2019semiparametric}. Product partition models separate the data into a set of contiguous clusters and evaluate the posterior probability as a product distribution \citep{monteiro2011product, park2010bayesian, garcia2019nonparametric}.  Many alternatives to MCMC sampling such as recursive Bayesian inversion paired with importance sampling \citep{Tan_2015}, simulation-based approaches \citep{wyse2010simulation} and conditional maximization \citep{fuentes2019modal} have also been considered. However, many of these approaches are unreliable in the presence of unlabeled outliers or heteroskedastic noise.

Outliers and heteroskedastic noise make changepoint estimation substantially more difficult by disproportionately impacting the distributional estimates of the data. Despite their common presence in many real world datasets, few changepoint algorithms have been able to account for them. For dealing with outliers, \citet{Fearnhead_2017} proposed an alternative biweight loss function to cap the influence of individual observations. While successful, the algorithm is sensitive to tuning parameters for both the influence threshold and the number of changepoints, which are difficult to specify in practice. \citet{pein2017heterogeneous} proposed an algorithm called H-SMUCE for identifying changepoints in presence of heterogeneous noise. H-SMUCE optimizes a constrained maximum likelihood function which minimizes the number of changepoints with the constraint that each segment's local log-likelihood ratio should be below a certain threshold. As we will show in the simulations, our algorithm out-performs both in presence of significant outliers or stochastic volatility. 

In this paper, we propose a new method called Adaptive Bayesian Changepoints with Outliers (ABCO). At its core, ABCO utilizes a threshold autoregressive stochastic volatility process within a time-varying parameter model to estimate smoothly varying trends with large, isolated interruptions. Through a state-space framework, ABCO successfully models heteroskedastic measurement error, and jointly identifies both local outliers and changepoints. Within the observation equation, we decompose a series into three basic components: a locally varying trend signal, a sparse additive outlier signal, and a heteroskedastic noise process. Our specification allows locally adaptive modeling for both trend and variability, automatically adjusting changepoint and outlier detection to periods of high and low volatility. 

ABCO's trend modeling greatly extends the Bayesian trend filter approach of \citet{kowal_2018} to include interruptions and the estimation of changepoint locations. Local smoothing is accomplished by modeling increments of the trend process as a sparse signals through global-local shrinkage priors \citep{Carvalho_2009}. To diminish insignificant patterns and merge sustained trends together, while still allowing instantaneous changes, we specify a local shrinkage `process' prior through a threshold stochastic volatility model with appropriately distributed innovations. A threshold parameter establishes a self-correcting mechanism that penalizes the occurrence of consecutive changepoints within short intervals. This also allows for posterior inference of changepoint locations and size. This is all performed while also accounting for additive outliers which we model with an ultra sparse horseshoe+ prior \citep{Bhadra_2017}. Finally, by combining locally estimated outlier and noise parameters we also define a posterior sample `local outlier score' for labeling each observation as likely anomalous or not.

The paper proceeds as follows. Section 2 details ABCO and our novel utilization of horseshoe priors. Section 3 contrasts ABCO with many alternative methods, including a basic horseshoe approach, dynamic shrinkage \citep{kowal_2018}, R-FPOP \citep{Fearnhead_2017}, E.Divisive \citep{Matteson_2013, Zhang_2019}, and H-SMUCE \citep{pein2017heterogeneous}, in diverse simulation scenarios. Section 4 investigates two real world applications using ABCO. Section 5 concludes with discussion of possible extensions.

\section{Methodology}
ABCO decomposes a time series $\{y_t\}$ as the sum of three components: a local mean or trend signal $\{\beta_t$\}, a sparse additive outlier signal $\{\zeta_t\}$, and a heteroskedastic noise process $\{\epsilon_t\}$. Specifically, we have
\begin{equation}\label{obs}
y_t = \beta_t + \zeta_t + \epsilon_t,  \quad \epsilon_t \sim N(0, \sigma_{\epsilon, t}^2).
\end{equation}
ABCO's decomposition effectively distinguishes a potentially locally varying mean or trend signal from a complex error process $\{\zeta_t + \epsilon_t\}$ that may contain outliers and exhibit non-constant volatility. Next, we detail 
each of the three model components of ABCO. 

\subsection{Trend Filtering with Changepoints}\label{trend}
The local mean or trend signal $\{\beta_t\}$ is specified as our primary state variable. To model a local trend, we focus on modeling increments in $\beta_t$, e.g., $\bigtriangleup^D \beta_{t}$, where $\bigtriangleup^D$ is the $D$th difference operator (usually for $D$ equal to 1 or 2), with global-local shrinkage priors. We suppose the following model
\begin{align}\label{state}
\bigtriangleup^D \beta_{t} &\equiv \omega_{t}, 
&\omega_t &\sim N(0, \tau_\omega^2\lambda_{\omega, t}^2), \nonumber \\
h_t &\equiv \log(\tau_\omega^2\lambda_{\omega, t}^2),
&h_{t} &= \mu+ (\phi_1 + \phi_2 s_{t})(h_{t-1} - \mu) + \eta_{t}.  
\end{align}
The trend increments, i.e.\ the evolution error  $\{\omega_t\}$, are modeled by a global-local shrinkage prior with parameters $\tau_\omega$ and $\{\lambda_{\omega,  t}\}$. The transformation \eqref{state} implies $\log(\tau^2_\omega) = \mu$ and $\log(\lambda_{\omega, t}^2) = (\phi_1 + \phi_2 s_{t})(h_{t-1} - \mu) + \eta_{t}$. The global parameter $\tau_\omega$ controls the global evolution error scale while the local parameters $\{\lambda_{\omega,  t}\}$ shrink evolution errors locally with respect to the time index $t$. The setup \eqref{state} induces local smoothing of the state variable, resulting in a relatively smooth underlying trend, $\{\beta_t\}$, while allowing for instantaneous jumps which will be classified as changepoints. The degree of differencing, $D$, can be interpreted roughly as modeling changes as locally linear ($D=1$) or locally quadratic ($D=2$) in trend; the former is more appropriate when the target has a piece-wise constant trend, for example. Ultimately, the degree of differencing is chosen by the user and depends on the use case and application.

The shrinkage process on the logarithmic scale, $\{h_t\}$, is modeled through a first order autoregression with $Z-$distributed innovations, $\eta_{t} \stackrel{iid}{\sim} Z(\alpha, \beta, 0, 1)$, where $Z(\cdot)$ denotes the four parameter $Z-$distribution with probability density function:
\begin{align*}
& p(z) = [B(\alpha, \beta)]^{-1}(\exp(z))^{\alpha} (1+\exp(z))^{-(\alpha + \beta)}, & z \in \mathbb{R}.
\end{align*} 
Previous work has shown a first-order stochastic volatility model for the process $\{\omega_t\}$ with $Z(\frac{1}{2}, \frac{1}{2}, 0, 1)$ distributed innovations have produced exceptionally flexible shrinkage appropriate for adaptive Bayesian trend filtering, i.e., $y_t = \beta_t + \epsilon_t$ \citep{kowal_2018}. Therefore, we fix the  $(\alpha, \beta)$ hyperparameters to $(\frac{1}{2}, \frac{1}{2})$ throughout.

The latent indicator $s_t$ and additional coefficient $\phi_2$ of \eqref{state} extends \citet{kowal_2018} to trend filtering with changepoint detection, but maintains intuitive simplicity. Our more flexible stochastic volatility model for $\{\omega_t\}$ is a first-order \emph{threshold} stochastic volatility TSV(1) model with shrinkage. This is equivalent to specifying a first order threshold autoregression for $\{h_t\}$ with $Z-$distributed innovations. The threshold indicator process $\{s_{t}\}$ of \eqref{state} induces an asymmetric response in the volatility model, $\phi_1$ versus $\phi_1 + \phi_2$, in $\{h_t\}$.  Specifically, for a threshold parameter $\gamma$, we define $s_t =  \textrm{I}( \log(\omega_{t-D}^2) > \gamma)$ where $ \textrm{I}(\cdot)$ is the indicator function. Time-steps in which $\log(\omega_t^2)$ exceed the threshold parameter will be classified as changepoints. For segments between changepoints, the posterior shrinkage profile for $\{\omega_{t}\}$ will tend to swing between persistent periods of shrinkage, in which $\omega_{t}$ is estimated as approximately zero, and periods of minimal shrinkage, in which $\omega_{t}$ is volatile and hence the trend $\beta_t$ itself is dynamically evolving. 

To motivate the inclusion of the threshold mechanism, consider a time series with the presence of changepoints in the signal $\beta_t$. The changepoints coincide with large values of $\log(\omega_t^2)$ and thus $h_t$ is inflated. When there is positive autocorrelation, $\phi_1$, in \eqref{state}, the subsequent process $\{h_t\}$ will remain non-zero for several periods because of the persistence induced by strong short-term memory of the autoregression. When now applying a TSV(1) prior, with the threshold indicator $s_t$ added, and with $\phi_2 < 0$, the process $\{ h_t \}$ can immediately return to a lower level following a large value of $\log(\omega_t^2)$. Overall, use of the proposed TSV(1) specification avoids over-estimation of changepoints especially in high volatility periods. 

Formally, the shrinkage properties can be assessed through the conditional posterior for the shrinkage proportion $\kappa_{t+1} = \frac{1}{1+\tau^2\lambda_{t+1}^2} \in [0, 1]$ under TSV(1) :
$$
[\kappa_{t+1} | y_{t+1}, \{\kappa_s\}_{s \le t}, s_t, \phi_1, \phi_2, \tau_\omega] = (1-\kappa_{t+1})^{-1/2}[1+(\psi_t-1)\kappa_{t+1}]^{-1}\exp(-y_{t+1}^2\kappa_{t+1}/2).
$$
where $\psi_t = (\tau_\omega^{2})^{(1-\phi_1-\phi_2s_t)}(\frac{1-\kappa_t}{\kappa_t})^{\phi_1+\phi_2 s_t}$. Assuming all other parameters are fixed, the posterior distribution of $\kappa_{t+1}$ conditional on $s_t = 1$ has more mass near 0 (unshrunk estimate) in comparison to the posterior distribution of $\kappa_{t+1}$ conditional on $s_t = 0$. This further highlights the purpose of the threshold variable; $\{s_t\}$ (and $\phi_2$) lowers the shrinkage value of $\kappa_{t+1}$ after the occurrence of a changepoint. In this way, ABCO separates isolated jumps in the trend from sustained periods of evolution in the trend.  \\
\pmb{Remark 1}. Let 
$\kappa_t = (1+\{\tau_{\omega}^2\lambda_{\omega, t}^2\})^{-1}$ denote the shrinkage proportion at time $t$, where as $\kappa_t \rightarrow 0$ there is no shrinkage and as $\kappa_t \rightarrow 1$ there is maximal shrinkage. Assume $y_t \sim_{ind} N(\omega_t, 1)$ and $\phi_1, \phi_2$ are fixed with $\phi_1 > \phi_2$. Let $\psi_t = (\tau_\omega^{2})^{(1-\phi_1-\phi_2s_t)}(\frac{1-\kappa_t}{\kappa_t})^{\phi_1+\phi_2 s_t}$, the following properties hold for the shrinkage of the trend component in ABCO:
\begin{enumerate}
    \item[(i)]  For any $\varepsilon \in (0, 1)$, $P(\kappa_{t+1} > 1-\varepsilon | y_{t+1}, \{\kappa_s\}_{s \le t}, s_t, \phi_1, \phi_2, \tau_\omega) \rightarrow 0$ as $\psi_t \rightarrow 0$ uniformly in $y_{t+1} \in \mathbb{R}$; 
    \item[(ii)] For any $\varepsilon \in (0, 1)$ and $\psi_t < 1$, $P(\kappa_{t+1} < \varepsilon | y_{t+1}, \{\kappa_s\}_{s \le t}, s_t, \phi_1, \phi_2, \tau_\omega) \rightarrow 1$ as \\ $|y_{t+1}| \rightarrow \infty$.
\end{enumerate}
The proof for both properties can be derived analogously to the proof of Theorem 3 in \citealt{kowal_2018}. The first property notes that ABCO will shrink the variance of the state equation toward 0 as $\tau_\omega \rightarrow 0$, confirming $\tau_\omega^2$ is a global shrinkage parameter in ABCO. The second property notes that a sufficiently extreme value of $y_{t+1}$ will still lead to a large change in the underlying mean trend, and this also motivates the use of the outlier process $\{ \zeta_t\}$ to redistribute the impact of such an extreme, particularly when it is isolated. 

\subsection{Local Outlier Detection and Outlier Scoring} \label{subsec_met_out}
The outlier process $\{\zeta_t\}$ models large deviations at specific times with an independent horseshoe+ shrinkage prior \citep{Bhadra_2017}. After conditioning on the data, the process $\{\zeta_t\}$ concentrates near zero except at locations where deviations from the mean trend are much larger in magnitude than nearby deviations. Thus, we call it a sparse outlier process. Although outliers may cluster, they are assumed \emph{a priori} independent and: 
\begin{align*}
    (\zeta_t | \lambda_{\zeta t}) &\sim N(0, \lambda_{\zeta, t}^2),
     &
    (\lambda_{\zeta, t} | \tau_\zeta, \eta_{\zeta, t}) &\sim C^+(0, \tau_\zeta \eta_{\zeta, t}),   \\
    \tau_\zeta &\sim C^+(0, \sigma_{\tau, \zeta}), 
     &
    \eta_{\zeta, t} &\sim C^+(0, \sigma_{\eta, \zeta}),
\end{align*}
where $C^+$ denotes the half-Cauchy distribution, and $\sigma_{\tau, \zeta}$ and $\sigma_{\eta, \zeta}$ are hyperparameters related to the global and local shrinkage of outliers, respectively.  With infrequent outliers, the outlier process $\{\zeta_t\}$ rarely takes values outside a tight neighborhood of zero. The horseshoe+ prior, with an additional variance term $\{\eta_{\zeta, t}\}$ compared to the horseshoe, provides more extreme horseshoe-shaped shrinkage (more prior mass at 0 and 1), allowing for large deviations from zero at only a small number of locations. 

In order to identify specific observations $y_t$ as potential outliers, we propose a locally adaptive outlier score $o_t$ based on the proportion of conditional variance attributed to the outlier component $\zeta_t$ relative to the variance of the overall error $\zeta_t + \epsilon_t$. From \eqref{obs} and the additional specification details discussed earlier, we note $\mathrm{Var}(y_t | \beta_t, \lambda_{\zeta, t}, \sigma_{\epsilon, t}) = \lambda_{\zeta, t}^2 + \sigma_{\epsilon, t}^2$, i.e., conditional on the local trend $\beta_t$, variability is split between the outlier and heteroskedastic noise terms. Thus, our proposed observational outlier score is 
$o_t = \textrm{E} \left[\frac{\lambda_{\zeta, t}^{2}}{\lambda_{\zeta, t}^{2} + \sigma_{\epsilon, t}^{2}}\right]$ where the expectation is with respect to the posterior. Outlier scoring provides a single ordering of the observations with respect to their relative local deviations. We shade the observation points by the outlier scores in subsequent figures. Thresholding the componentwise outlier scores is a simple approach to labeling locally outlying points and comprehensive joint analysis of outlier scores is a promising future research direction.

\subsection{Heteroskedastic Noise Process}
We assume the noise process  $\{\sigma_{\epsilon, t}^2\}$ follows a stochastic volatility (SV) model to capture potential heteroskedasticity in the observations. We focus on the first order SV(1) model of \citep{kim_1998} with Gaussian innovations in log volatility. This relatively simple noise model is robust and widely adaptive in practice; more complex specifications could be substituted into ABCO for specific applications. It is important to note that the model does not require the true noise to be stochastic; rather, this noise process can deal well with both homoskedastic and heteroskedastic data (see Figure \ref{fig2} for an illustration). Specifically, the model can be written as
\begin{equation*}
\log(\sigma_{\epsilon, t}^2) = \mu_\epsilon + \phi_\epsilon[ \log(\sigma_{\epsilon, t-1}^2)-\mu_\epsilon]+\nu_{\epsilon, t}, \quad \nu_{\epsilon, t} \sim N(0, \sigma_{\nu}^2).
\end{equation*}

The SV(1) model concisely accounts for heteroskedastic measurement error. We also compare to a model with constant measurement error variance by supposing $\sigma_{\epsilon, t}^2$ is a constant, $\sigma_\epsilon^2$, for all $t$ which we later denote as `ABCO w/o SV'. As shown in Section \ref{sec_sim}, incorporating stochastic volatility greatly improves changepoint detection in series with heteroskedastic noise, while maintaining an equal level of performance in series with constant noise variance at minimal computational cost. This simple SV(1) process has not been widely adopted in other changepoint models and further highlights ABCO's enhanced flexibility for modeling  more varied and complex applications. 

Overall, the three components of ABCO ($\{\beta_t\}, \{\zeta_t\}, \{\epsilon_t\}$) make up a robust and flexible model for detecting changepoints in noisy time series. The full details of the posterior sampling process are given in the next section. 

\section{Estimation and Inference}\label{sec:samp}

We conduct posterior inference via Markov chain Markov Carlo. In particular, the algorithm used Gibbs sampling when full conditional distributions could be derived. In this section, we provide an overview of the sampling steps of ABCO. Note that some of these samplers follow closely from related works \citep{kowal_2018, Kastner_2014}. The exception is the threshold variable $\gamma$, which is more difficult to estimate and sampling details are summarized below.

Let $\pmb{Y} = [y_1, ..., y_{T}]$, $\pmb{h} = [h_1, ..., h_{T}]$, $\pmb{\eta} = [\eta_1, ..., \eta_{T}]$, $\pmb{\beta} = [\beta_1, ..., \beta_T]$, $\pmb{\sigma}_\epsilon^2 = [\sigma_{\epsilon, 1}^2,..., \sigma_{\epsilon, T}^2]$, $\pmb{\zeta} = [\zeta_1, ..., \zeta_T]$, $\pmb{\lambda}_\zeta = [\lambda_{\zeta, 1}, ..., \lambda_{\zeta, T}]$, $\pmb{\eta}_\zeta = [\eta_{\zeta, 1}, ..., \lambda_{\eta, T}]$ and $\pmb{\nu}_\epsilon = [\nu_{\epsilon, 1}, ..., \nu_{\epsilon, T}]$. 
Sampling all variables in the evolutionary equation \eqref{state} from their full conditional
involves sampling the log volatility, $\pmb{h}$, the unconditional mean $\mu$, the evolution equation coefficients, $\phi_1$, $\phi_2$, the threshold, $\gamma$, and the evolution error, $\pmb{\eta}$. The state variable is sampled jointly from $p(\pmb{\beta}| \pmb{h}, \phi_1, \phi_2, \mu, \gamma, \pmb{\eta}, \pmb{\sigma}_\epsilon^2, \pmb{\zeta}, \pmb{Y})$. The outlier process is sampled from $p(\pmb{\zeta} | \pmb{\beta}, \pmb{\sigma}_\epsilon^2, \pmb{\lambda}_\zeta, \tau_\zeta, \pmb{\eta}_{\zeta}, \pmb{Y})$ along with other associated parameters using a sampler for the horseshoe prior based on the inverse gamma distribution \citep{Makalic_2016}. And the observational variance is sampled from $p(\pmb{\sigma}_\epsilon^2 | \pmb{\beta}, \pmb{\zeta}, \mu_\epsilon, \phi_\epsilon, \pmb{\nu}_\epsilon, \pmb{Y})$ along with other associated variables using the stochvol R package \citep{Hosszejni_2019}. From those samples, the increments $\{\omega_t\}$, global shrinkage $\tau_\omega^2$, local shrinkage $\{\lambda_t^2\}$, and changepoint indicators $\{s_t\}$ can be calculated from the stated relationships.

\subsection{Sampling $\gamma$}
The update for $\gamma$ is less straightforward relative to the other parameters above. Following derivations in \cite{Nakajima_2013}, for each time step $t$, marginalizing over $\omega_t$ gives $P(\log(\omega_t^2) > \gamma) = P(\omega_t > e^{\gamma/2}) = 1 - \Phi(\frac{e^{y_t/2}}{\tau\lambda_t})$ where $\Phi$ is the CDF for standard normal distribution. From this we see that $\tau$, $\{\lambda_t\}$ and $\{y_t\}$ play important roles in identifying an appropriate range for $\gamma$. However, the full conditional distribution for $\gamma$ does not have a standard form. Despite this challenge, it is important to learn $\gamma$ from the data, even in the presence of outliers, because it plays a crucial role in determining the location of the changepoints. After experimenting with numerous sampling schemes, including slice sampling \citep{Neil_2003} and a Griddy Gibbs sampler \citep{Ritter_1992}, we find that applying Metropolis-Hastings within a Gibbs sampler \citep{Nakajima_2013} was both simple and adequate for estimating $\gamma$ within the ABCO model and allows essentially any choice of prior. We further found that a uniform prior, $\gamma \sim \text{Unif}(\ell_\gamma, u_\gamma)$, worked well provided the range was tailored to the observed data, and we relate the lower and upper limits to the volatility of the $D$th degree differences of the observations. Specifically, we let $\ell_\gamma = \min\{\log[(\bigtriangleup^D y_t)^2]\}$ and $u_\gamma = Q_{q}\{\log[(\bigtriangleup^D y_t)^2]\}$ where $Q(\cdot)$ is the quantile function and $q$ is a fixed hyper-parameter specifying the choice of quantile. For most settings, choosing $q=1$ or the highest quantile produces a reasonable upper bound. However, for problems in which the shifts in signal is significantly smaller than the observational variance.

\subsection{Sampling the log evolution variance $\{h_t\}$}
To sample the log evolution variance, we will use a similar sampling method as detailed in \cite{Kastner_2014}. In details below, we will describe the setup for $D=1$; the sampler for $D=2$ follows similarly. 

The evolutionary log variance equation is given by $h_{t+1} = \mu + (\phi_1 + \phi_2 s_{t})(h_t - \mu) + \eta_{t+1}$. To sample the log-volatility $\{h_t\}$, we will first define $z_t = log(\omega_t^2+c)$. Since each $\omega_t$ follows a Gaussian distribution conditionally, each $\omega_t^2$ follows a chi-squared distribution, conditionally. To sample the log of a chi-squared distribution, we use the 10-component discrete mixture approximation with mean, variance and weights denoted by $m_{i}, v_{i}, w_{i}$ for $i \in \{1,..,10\}$ \citep{omori_2007}. The discrete mixture indicators $r_t$ are sampled in the same manner as in \cite{kim_1998}. The joint likelihood for all $h_1, ..., h_T$ as follows: $\pmb{z} \sim N(\pmb{m + \tilde{h} + u}, \pmb{\Sigma}_{v})$ where $\pmb{z} = (z_1, ..., z_T)$, $\pmb{m} = (m_{r_1}, ..., m_{r_T})$, $\pmb{\tilde{h}} = (h_1 -\mu, ..., h_T - \mu)$, $\pmb{\mu} = (\mu, \mu ..., \mu)$, and $\pmb{\Sigma}_{v} = diag(v_{r_1}, ... v_{r_T})$. The evolution equation can be written as $\pmb{D \tilde{h}} \sim N(\pmb{0},\pmb{\Sigma}_\nu)$ where $\pmb{\Sigma}_\nu = diag(\nu_1, ..., \nu_T)$ and $\{\nu_t\}$ are the  P\'{o}lya-Gamma auxiliary variables of the sampling for  $\{\eta_t\}$ described in the next subsection. $\pmb{D}$ is a T $\times$ T matrix with 1 on the diagonal, $(-(\phi_1+\phi_2s_2), ..., -(\phi_1+\phi_2s_{T}))$ on the first off-diagonal and 0 elsewhere. 

With the above setup, we can calculate the parameters of the full conditional distribution for $\pmb{\tilde{h}}$ using the \textit{all without a loop sampler} detailed in \cite{Mccausland_2011}. The full conditional distribution is given by:  $\pmb{\tilde{h}} \sim N(\pmb{Q}^{-1}\pmb{l}, \pmb{Q}^{-1})$ where $\pmb{Q}$ is a symmetric tridiagonal matrix with main diagonal entries $\pmb{Q_0}$ and off-diagonal entries $\pmb{Q_1}$ given by
\begin{align*}
\pmb{Q_0} &= [(v_{r_1}^{-1}+\nu_1+(\phi_1+\phi_2s_2)^2\nu_2),...,(v_{r_{T-1}}^{-1}+\nu_{T-1}+(\phi_1+\phi_2s_T)^2\nu_T), (v_{r_T}^{-1} + \nu_T)] \\
\pmb{Q_1} &= [-((\phi_1+\phi_2s_2)\nu_2), -((\phi_1+\phi_2s_3)\nu_3)..., -((\phi_1+\phi_2s_{T})\nu_{T})] \\
\pmb{l} &= [\frac{z-m_{r_1}-\mu}{v_{r_1}}, \frac{z-m_{r_2}-\mu}{v_{r_2}}, ..., \frac{z-m_{r_T}-\mu}{v_{r_T}}]
\end{align*}
With the full conditional distribution derived above, we sample $\pmb{\tilde{h}}$ using the \textit{backhand-substitution method} detailed in \cite{Kastner_2014}.

\subsection{Sampling the remaining parameters}
Since a horseshoe prior is used for the evolution error,we propose the following prior for $\tau_\omega$: $\tau_\omega \sim C^+(0, \frac{1}{\sqrt{T}})$. Using P\'{o}lya-Gamma mixing parameter as seen in \cite{Polson_2013}, the prior for $\mu$ is given by: $[\mu |\tau] \sim N(\log(\frac{1}{T}), \nu_\mu^{-1})$, where $\nu_\mu \sim PG(0, 1)$. Given that $h_1 \sim N(\mu, \nu_0^{-1})$ where $\nu_0 \sim PG(0, 1)$, the full conditional distribution for $\mu$ can be derived from formula for conjugate normal prior. The full conditional distribution is given by: $\mu \sim N(Q_\mu^{-1}l_\mu, Q_\mu^{-1})$, where
\begin{align*}
Q_\mu &= \nu_\mu+\nu_0 + \sum_{t=1}^{T-1}(1-\phi_1-\phi_2s_{t+1})\nu_t, \\
l_\mu &= \nu_\mu log(\frac{1}{T}) + \nu_0 h_1 + \sum_{t=1}^{T-1}(1-\phi_1-\phi_2s_{t+1})(h_{t+1}-(\phi_1+\phi_2s_{t+1})h_t) \nu_t.
\end{align*}

For $\phi_1$, assume a prior $\frac{\phi_1+1}{2} \sim \text{Beta}(20, 1)$, which restricts $|\phi_1| \le 1$. The full conditional distribution is then sampled using a slice sampler with a lower limit of 0 and a upper limit of 1 \citep{Neil_2003}. Slice sampling provides an effective way to sample a univariate distribution with known lower and upper limit. The full conditional is given by: 
\begin{align*}
f_{(\phi_1+1)/2}(x) = [-\frac{1}{2} \sum_{t=1}^{T-1} (\tilde{h}_{t+1} - (2x - 1 + \phi_2s_{t+1})\tilde{h}_t)^2] + \text{dBeta}(x, 20, 1)
\end{align*}
where $\text{dBeta}(x, 10, 2)$ is the pdf of the beta distribution with parameters $\alpha = 10$ and $\beta = 2$. 

For $\phi_2$, we assume a prior truncated normal distribution with lower bound of 0, $\phi_2 \sim N(-1, 0.5)\textbf{1}_{\phi_2 \le 0}$. We restrict $\phi_2$ to be negative as $\phi_2$ functions as a negative buffer after discovering a changepoint. The full conditional can be calculated by defining $\pmb{s} = \{t; s_{t+1} = 1\}$; it is given by:
\begin{align*}
f_{\phi_2}(x) = [-\frac{1}{2} \sum_{t \in \pmb{s}} (\tilde{h}_{t+1} - (\phi_1 + xs_{t+1})\tilde{h}_{t})^2] + \text{dNorm}(x, -1, 0.5)
\end{align*}
where $\text{dNorm}(x, -1, 0.5)$ is the pdf of the normal distribution with mean $-1$ and standard deviation of $0.5$. The full conditional distribution is sampled with lower limit of -5 and upper limit of 0. 

The evolution error $\{\eta_t\}$ is defined by a P\'{o}lya-Gamma hierarchical parameterization (\cite{Polson_2013}) given by $[\eta_t|\nu_t] \stackrel{ind.}{\sim}  N(0, \nu_t^{-1})$ where $\nu_t^{-1} \sim PG(1,0)$. The P\'{o}lya-Gamma parameters, $\{\nu_t\}$, are sampled using the rpg() function from Bayeslogic R package.  Next, the state variable $\{\beta_t\}$ is sampled in a similar manner as in \cite{kowal_2018}. Lastly, the observation variance $\epsilon_t$ follow a standard stochastic volatility model. The observational variances are sampled using stochvol R package.

\section{Simulation Experiments} \label{sec_sim}

In this section, we will show the effectiveness of ABCO in several challenging settings involving outliers and heteroskedasticity. We will compare against five other method: the independent horseshoe, dynamic shrinkage process of \citep{kowal_2018}, Robust FPOP algorithm \citep[R-FPOP,][]{Fearnhead_2017}, E.Divisive \citep{James_2014} and heterogeneous simultaneous multiscale change-point estimator \citep[H-SMUCE,][]{pein2017heterogeneous}. 

Both the horseshoe prior and dynamic shrinkage process assume a trend equation of $y_t = \beta_t + \epsilon_t$ with observation stochastic volatiliy. Additionaly, the horseshoe approach fixes $\phi_1, \phi_2$ at 0 from Equation (\ref{state}). Fixing the coefficients to be 0 leads to non-dynamic shrinkage, in that $\{\lambda_{\omega, t}\}$ are independent and identically distributed. The dynamic shrinkage process (DSP) specifies $\phi_2$ fixed at 0 and estimates $\phi_1$. The horseshoe and DSP function as baseline benchmarks against the more complicated evolution and trend processes of ABCO. Throughout, we take the posterior mean for the Bayesian point estimates and the 95\% highest posterior density (HPD) for the Bayesian interval estimates. 

R-FPOP utilizes a penalized likelihood estimator with a biweight loss to cap the impact of individual points. The approach is designed to identify changepoints in presence of outliers. E.Divisive is a non-parametric approach based on energy distance to identify optimal partitioning within a time series. H-SMUCE estimates a piece-wise constant function for the underlying data using a heterogeneous noise process. We chose these competitors as they represent a diverse set of robust changepoint algorithms. In particular, R-FPOP is designed to work in presence of outliers and H-SMUCE is designed to work in presence of heterogeneity. As a result, they make good benchmarks for comparison in these settings.  R-FPOP, E.Divisive and H-SMUCE are specified using default parameters. 

Four key metrics are used for comparison of results: Rand average, adjusted Rand average, F1-score and average distance to closest true changepoint. Rand Average measures similarity between predicted partition and true partition; the value ranges between 0 and 1, with 1 being a perfect match. Adjusted Rand average corrects Rand average for random chance of predicting the correct segmentation. F1-score measures the harmonic mean between precision and recall in the predicted changepoints. We classify a predicted changepoint as a true positive if it's within 5 time-steps of a true changepoint with the caveat that each true changepoint can match with at most one predicted changepoint. Average distances to closest true changepoint measures the average number of time-steps from predicted changepoints to the nearest true changepoint. 

\subsection{Multiple Changes in Mean with Heteroskedastic Noise} \label{subsec_sim_sv}

\begin{figure}[ht!]
  \centering
  \caption{Example Plots for Changes in Mean with Stochastic Volatility}
  \label{fig1} 
  \hspace*{0.01cm} 
  \includegraphics[width = \textwidth]{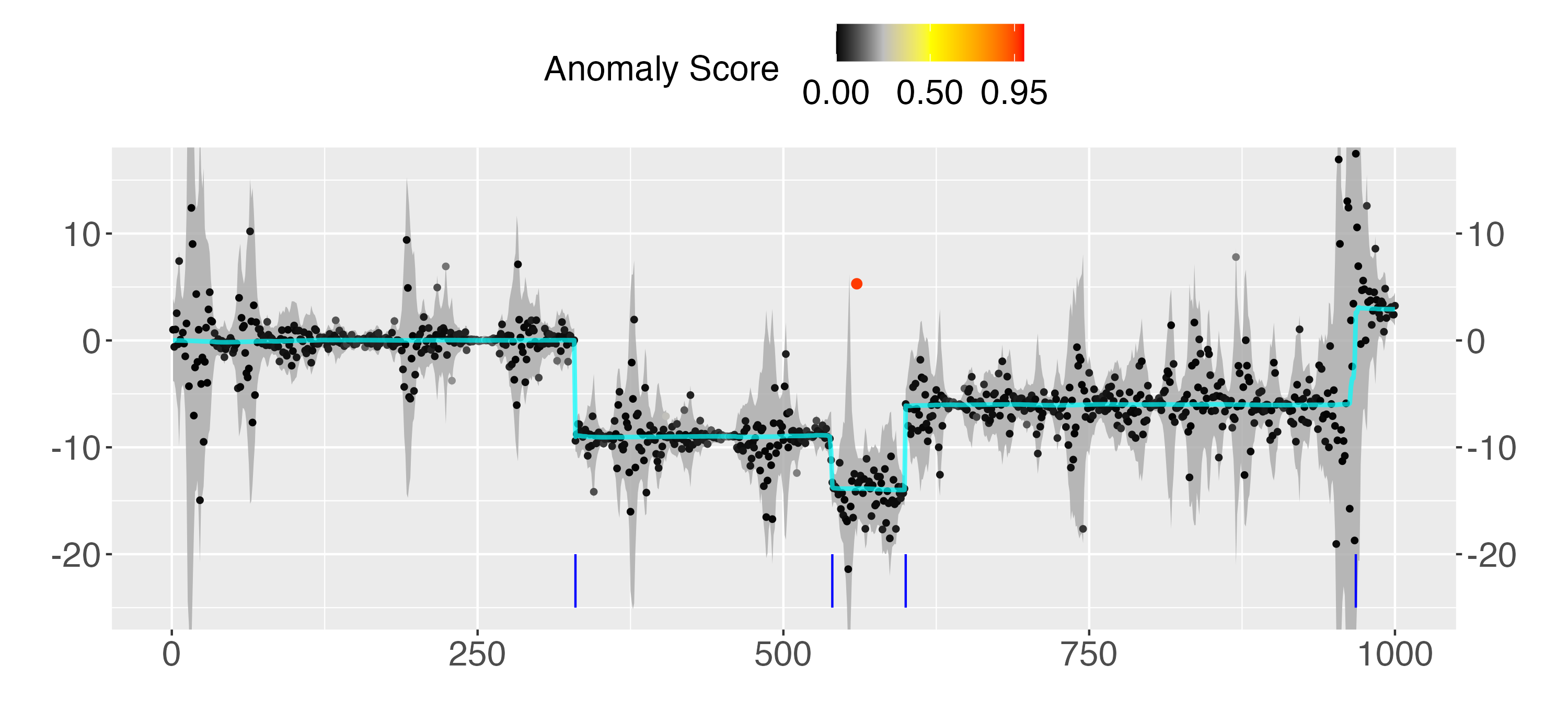} \\
  \includegraphics[width = 0.45\textwidth]{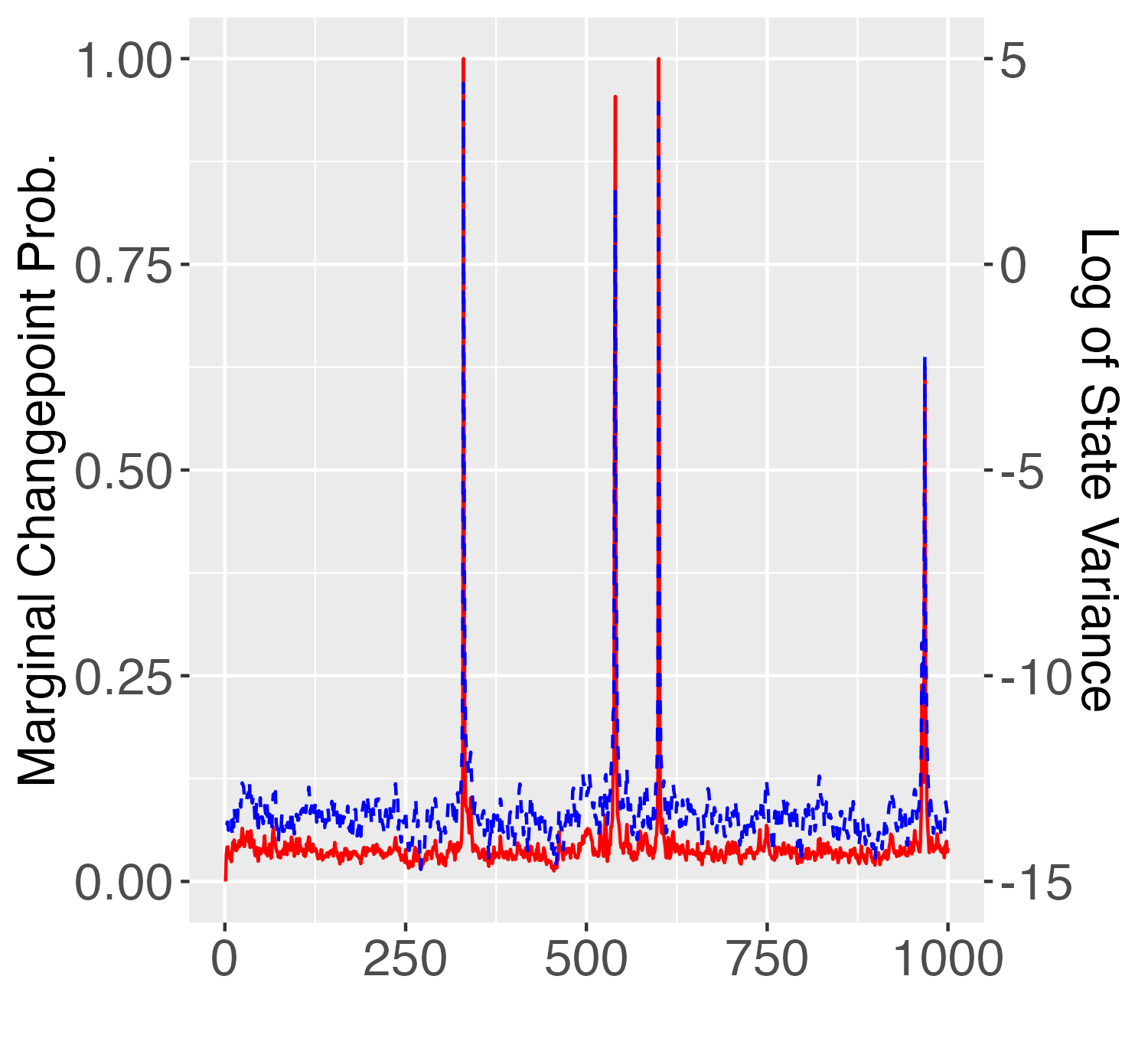}
  \includegraphics[width = 0.45\textwidth]{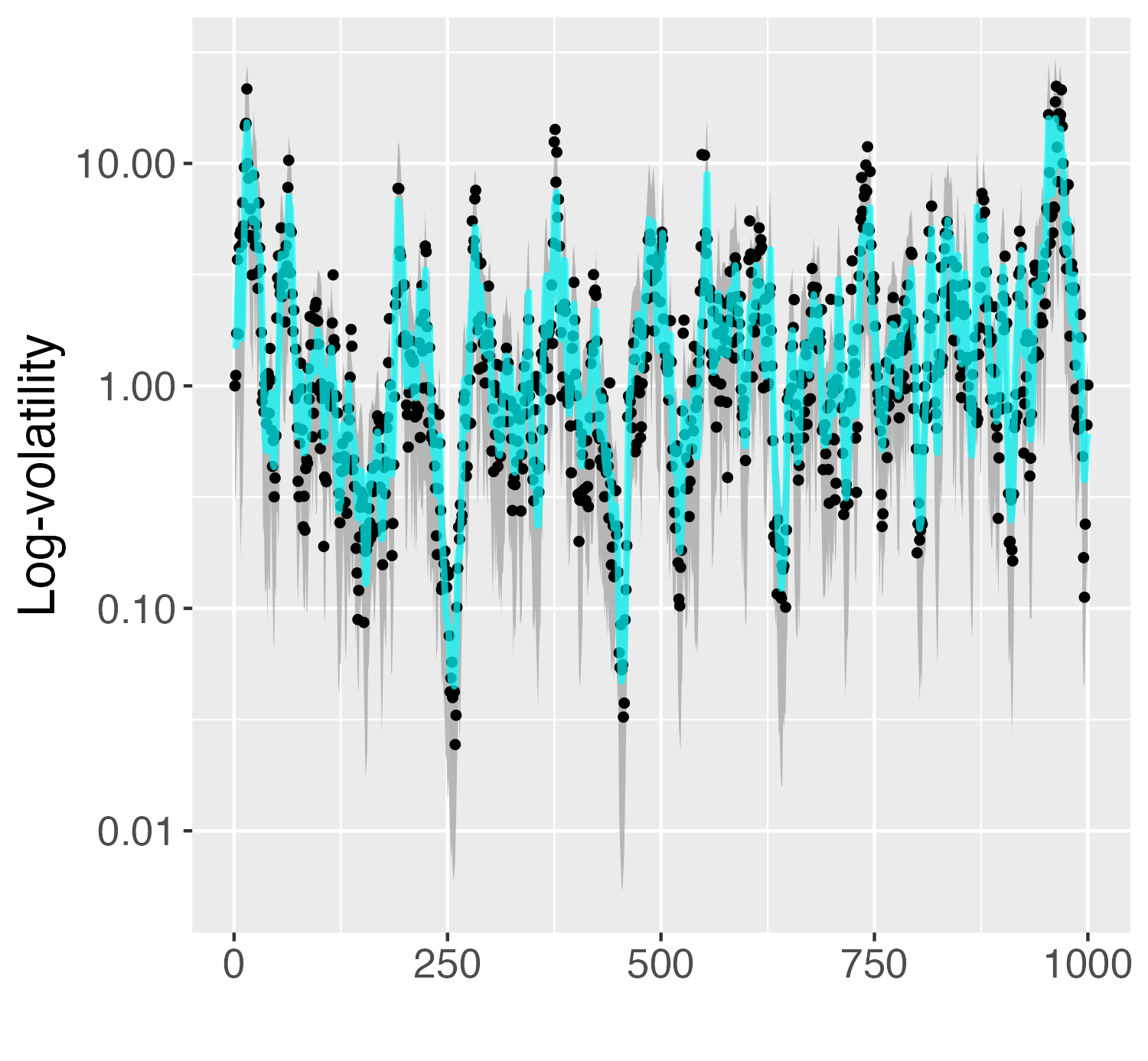}\\
  \begin{flushleft}
  \setlength{\baselineskip}{1.0pt}
{The top plot shows an example of simulated data with added outliers. The anomaly scoring shows local adaptive of the algorithm. The cyan lines indicate posterior mean of $\{\beta_t\}$; vertical blue lines indicate predicted changepoints and gray bands indicate $95\%$ point-wise credible bands for the data excluding the anomaly process component. The bottom plot shows the marginal probability of being a changepoint at each time step (red line) and $\{\log(\omega_{t-D}^2)\}$ (blue dashed line); the marginal probability at each time step $t$ is calculated from the percentage of posterior simulations of $\log(\omega_{t-D}^2)$ that exceed the changepoint threshold $\gamma$. We can see, as a result of the threshold variable, $\{\log(\omega_{t-D}^2)\}$ spikes when changepoint is predicted and comes down right-away; this is key for not over-predicting in regions of high-volatility. Lastly, we plot the posterior mean of $\{\sigma_{\epsilon, t}\}$ (cyan line) with 95\% point-wise credible intervals and true simulated values (black dots).}
\end{flushleft}
\end{figure}

First, we consider the effectiveness of ABCO in the setting of estimating multiple changepoints in mean in the presence of heteroskedastic noise. In this section we generate $N=100$ simulated series of length $T=1000$. For each series, the number of changepoints are generated uniformly at random from the set $\{2, 3, 4\}$. Changepoint locations are uniformly sampled but with the constraint that their minimal distance is 5 time increments to ensure identifiability. The resulting segments are each assigned a mean uniformly distributed between $-10$ and $10$. Finally, observation innovations with stochastic volatility are added to the mean signals, with log volatility specified as 
\begin{equation}
\log(\sigma_{\epsilon, t}^2) = \phi_\epsilon \log(\sigma_{\epsilon, t-1}^2) + \nu_{\epsilon, t},  \quad \nu_{\epsilon, t} \sim N(0, \sigma_\nu^2),
\end{equation}
with $\phi_\epsilon = 0.9$ and $\sigma_\nu^2 = 1$. More simulation results for varying signal-to-noise ratios are shown in the Online Appendix. Figure \ref{fig1} gives an example of such a generated series, but with two outliers also included for illustration here (outliers are considered further in Section \ref{subsec_sim_out} and \ref{subsec_sim_lin}). Each series tends to have substantial local fluctuations which make changepoint analysis difficult. 

\begin{table*}
\caption{Mean Changes with Stochastic Volatility}
\label{tab1}
\centering
\begin{tabular}{ c c c c c }
\hline \hline
  Algorithms & Rand Avg. & Adj. Rand Avg. & F1-Score & Avg. Dist. to True \\ 
 \hline
 ABCO  & $\textbf{0.95}_{(0.11)}$ & $\textbf{0.90}_{(0.21)}$ & $\textbf{0.88}$ & $\textbf{0.44}_{(0.80)}$ \\
 \hline
 DSP & $0.69_{(0.15)}$ & $0.41_{(0.26)}$ & $0.13$ & $130.31_{(120.40)}$ \\
 \hline
 Horseshoe & $0.68_{(0.16)}$ & $0.38_{(0.26)}$ & $0.21$ & $116.39_{(125.53)}$ \\
 \hline
 E.Divisive & $0.83_{(0.11)}$ & $0.63_{(0.21)}$ & $0.51$ & $73.41_{(42.31)}$ \\
 \hline
 R-FPOP  & $0.92_{(0.13)}$ & $0.85_{(0.24)}$ & $0.77$ & $18.97_{(59.36)}$ \\
 \hline 
 H-SMUCE & $\textbf{0.95}_{(0.10)}$ & $0.89_{(0.19)}$ & $0.76$ & $14.44_{(37.38)}$ \\
 \hline
\end{tabular}
\begin{flushleft}
  \setlength{\baselineskip}{1.0pt}
{Table \ref{tab1} details comparisons between ABCO, the horseshoe approach, R-FPOP, E.Divisive and H-SMUCE on simulated data with stochastic volatility. The value corresponds to summaries across fits to 100 simulated datasets and the standard deviations for the metrics are shown in subscript parentheses.}
\end{flushleft}
\end{table*}

Table \ref{tab1} details results comparing the performance of ABCO with the competing methods. As seen, ABCO performs the best among the competing models, highlighting its robustness to heteroskedastic noise. ABCO and H-SMUCE both achieve the highest average Rand value of 0.95 with standard errors of 0.11 and 0.10, respectively. ABCO achives the highest average adjusted Rand (0.90) and H-SMUCE is a close second (0.89). The competitiveness is expected as H-SMUCE is designed to perform well for data characterized by heteroskedastic noise. ABCO achieves a better F1-score of 0.88, in comparison to 0.76 of H-SMUCE, showing a better trade-off in terms of precision and recall for ABCO. R-FPOP achieves the second highest F1-score of 0.77 but has a lower average adjusted Rand value of 0.85. The DSP and horseshoe perform similar and uniformly poor. 

A further statistic to highlight is the average distance from a predicted changepoint to a true changepoint. ABCO has an impressive distance of 0.44 (se 0.80), with almost every predicted changepoint very close to a true changepoint. While R-FPOP and H-SMUCE were similar in terms of other metrics, they feature much greater prediction-to-true distances of 18.97 and 14.44 respectively. This implies that ABCO is more accurate at predicting the correct location and less susceptible to predicting false positives. The ability to avoid false positives in this setting makes ABCO well suited for diverse applications. In comparison, the static horseshoe and DSP are less accurate in predicting the correct number of changepoints and tends to miss some more subtle change as illustrated by low F1-scores of 0.21 and 0.13, respectively.  Although DSP and horseshoe are both designed to account for stochastic volatility, it appears the evolution process is too limiting.

Furthermore, within the Bayesian framework, ABCO estimates both the local mean and point-wise credible bands. This allows tracking both a constant or varying signal within each cluster. We can see in Figure \ref{fig1} that the credible bands are able to adapt well to regions of high volatility. Also, the true observation volatility, $\{\sigma_{\epsilon, t}\}$, is captured well by the model as shown in the bottom right panel of Figure \ref{fig1}.

\subsection{Detecting Changes in the Presence of Outliers} \label{subsec_sim_out}

Next we illustrate the effectiveness of ABCO in the presence of significant outliers. 100 time series are generated of length 300, with a single changepoint in mean randomly selected in the middle 50 percent of the series, partitioning the series into two segments. The first segment has a mean of 0 and the second segment has a mean of 2. Both segments have noise simulated by a normal distribution with mean 0 and variance 1. Additionally, for each segment, 5 significant outliers are generated uniformly at random to be 20-30 standard deviations away from the mean. The outliers are chosen to be significantly greater than the change in mean to test robustness against extreme values. We again fit the ABCO model, DSP, the static horseshoe model, E.Divisive, R-FPOP and H-SMUCE on the simulated data sets. 

\begin{table*} [ht!]
\caption{Mean Changes with Outliers}
\label{tab2}
\centering
\begin{tabular}{ c c c c c}
\hline \hline
 Algorithms & Rand Avg. & Adj. Rand Avg. & F1-Score & Avg. Dist. to True\\ 
 \hline
 ABCO & $\textbf{0.95}_{(0.13)}$ & $\textbf{0.91}_{(0.27)}$ & $\textbf{0.78}$ & $\textbf{1.32}_{(4.07)}$ \\
 \hline
 DSP & $0.66_{(0.08)}$ & $0.35_{(0.14)}$ & $0.02$ & $62.83_{(19.23)}$ \\
 \hline
 Horseshoe & $0.70_{(0.09)}$ & $0.42_{(0.18)}$ & $0.01$ & $63.10_{(23.80)}$ \\
 \hline
 E.Divisive & $0.84_{(0.09)}$ & $0.69_{(0.16)}$ & $0.40$ & $32.70_{(13.86)}$ \\
 \hline
 R-FPOP & $0.73_{(0.13)}$ & $0.45_{(0.26)}$ & $0.02$ & $55.76_{(34.31)}$  \\
 \hline
 H-SMUCE & $0.92_{(0.05)}$ & $0.84_{(0.11)}$ & $0.17$ & $12.99_{(10.39)}$ \\
 \hline
\end{tabular}
\begin{flushleft}
  \setlength{\baselineskip}{1.0pt}
{Result for ABCO against other changepoint algorithms on simulated data with changes in mean and significant outliers. The value corresponds to summaries across fits to 100 simulated datasets and the standard deviations for the metrics are shown in subscript parentheses.}
\end{flushleft}
\end{table*}

\begin{figure} [ht!]
\setlength{\tabcolsep}{2pt}
  \centering
  \caption{ABCO Outlier Scoring / Stochastic Volatility}
  \label{fig2}
   \begin{tabular}{c c}
  a) \includegraphics[scale=.095]{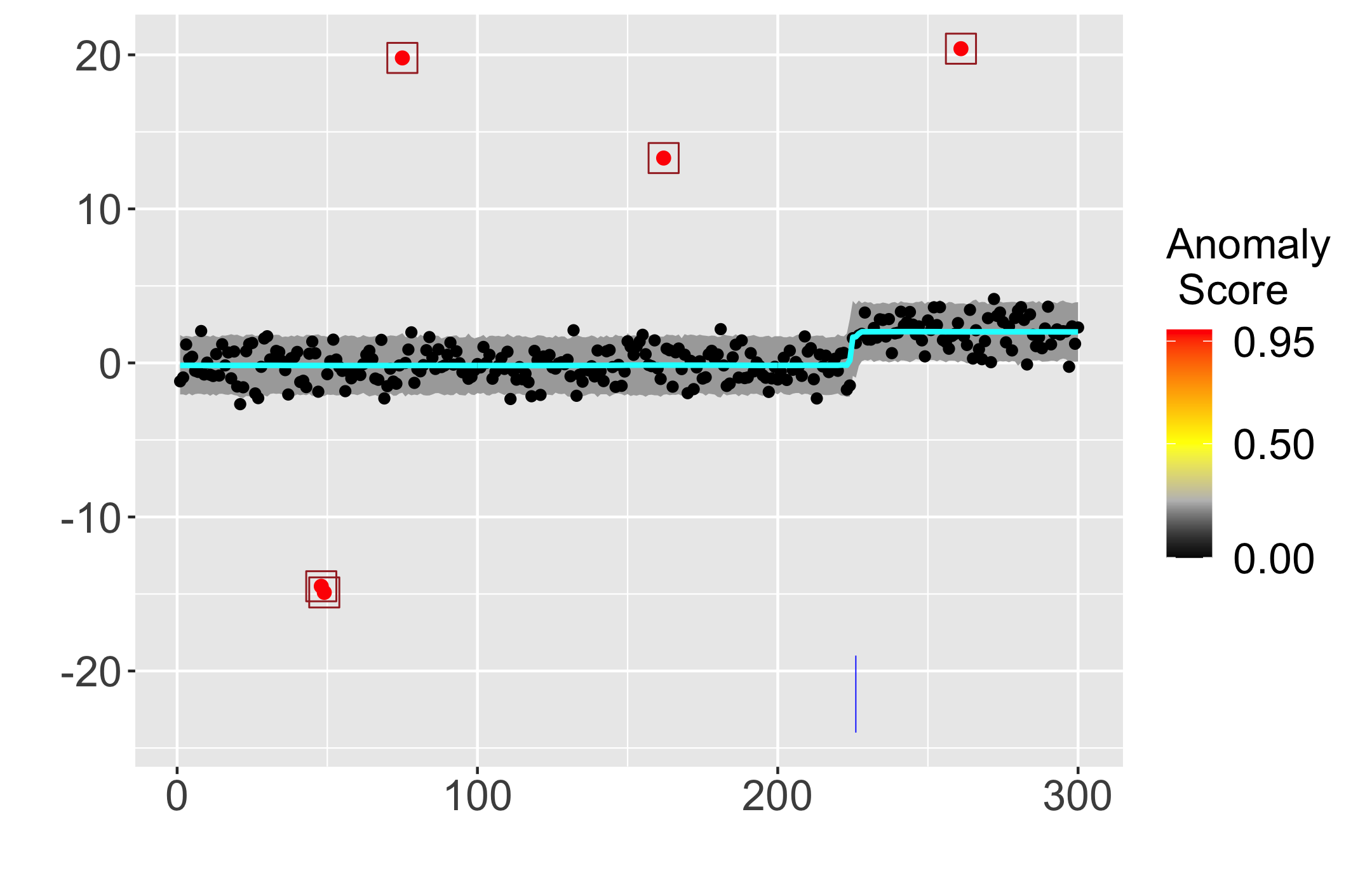} & c) \includegraphics[scale=.095]{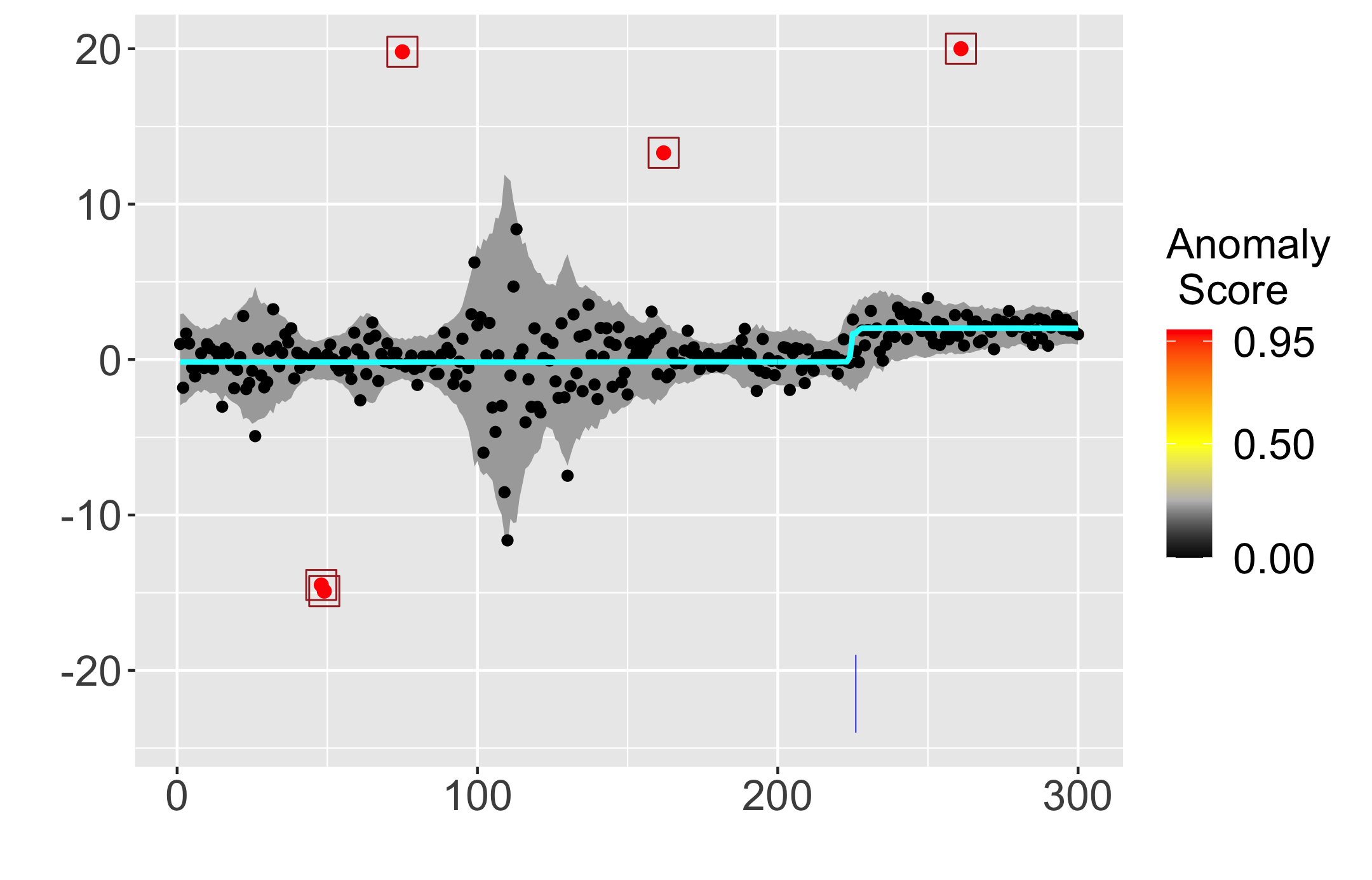} \\
  b) \includegraphics[scale=.095]{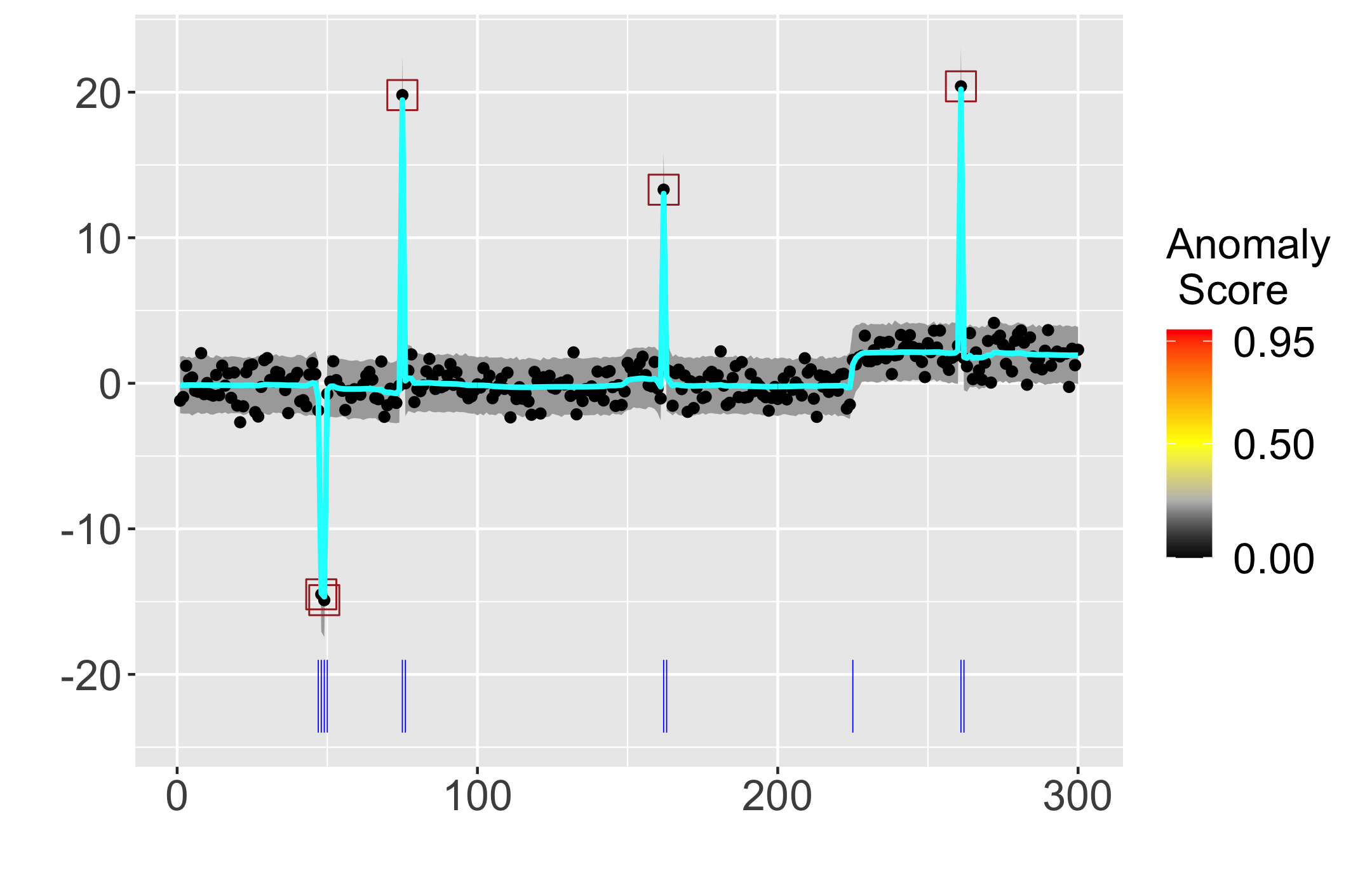} &
  d) \includegraphics[scale=.095]{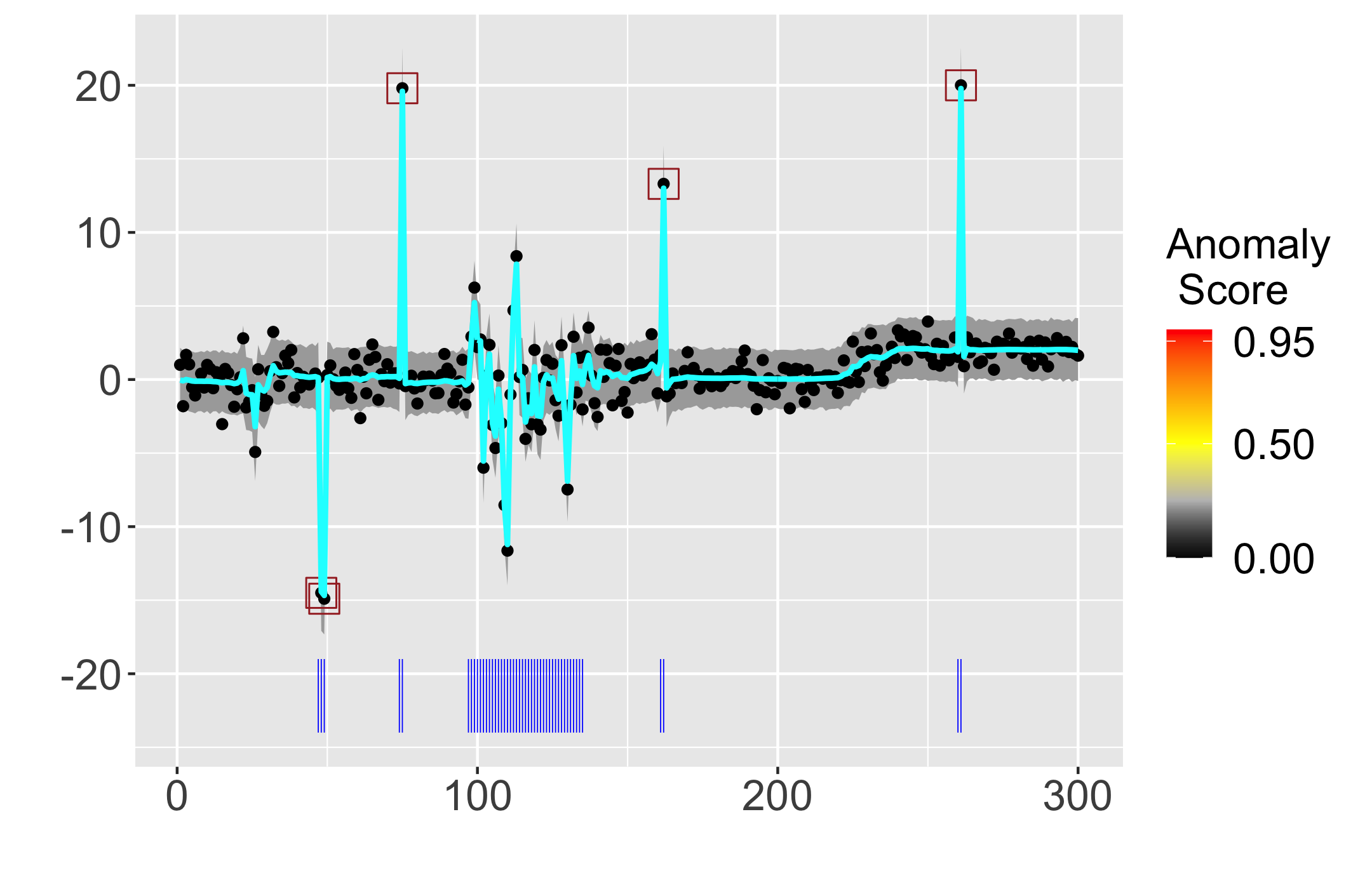} 
  \end{tabular}
  \begin{flushleft}
  \setlength{\baselineskip}{1.0pt}
   {This figure illustrates data generated with same signal and outlier but different noise processes. a) Results of ABCO on data generated with constant noise. b) Results of ABCO without the stochastic volatility component and without the anomaly component on same data as plot a. c) Results of ABCO on data generated with stochastic volatility. d) Results of ABCO without the stochastic volatility component and without the anomaly component on same data as plot c. As seen, ABCO struggles to adapt to these series without both the outlier and the stochastic volatility components included. True positive outliers flagged by the model are outlined by squares. The estimated mean (trend) is shown in cyan; estimated changepoint locations are indicated by blue vertical lines below the series, and credible bands for the observations excluding the estimated outlier process is shown in gray. }
   \end{flushleft}
\end{figure} 

The result from the simulation scenario is summarized in Table \ref{tab2}, where ABCO outperforms all other models. ABCO achieves highest average Rand value of 0.95 (se 0.13) and highest average adjusted Rand value of 0.91 (se 0.27). H-SMUCE came second overall with average adjusted rand of 0.84 (se 0.11), while all other algorithms performed quite a bit worse. The key difference between ABCO and H-SMUCE in that ABCO is more accurate at identifying the correct changepoint locations; this is reflected by an average distance from predicted to true changepoints of 1.32 for ABCO in comparison to 12.99 for H-SMUCE. As a result, ABCO has a much higher F1-Score than H-SMUCE. R-FPOP, an algorithm designed to work in presence of outliers, did not perform competitively against ABCO. R-FPOP sometimes will under-predict number of changepoints, leading to a much lower average adjusted Rand average and larger average distance to the true changepoint. Lastly, similar to results from Subsection \ref{subsec_sim_sv}, the static horseshoe prior and DSP significantly underperform compared to the evolution process of ABCO. 

As seen, ABCO works well in both settings of stochastic volatility and significant outliers. This is due to the fact we have decomposed the noise into two components: outlier and heteroskedastic noise. Figure \ref{fig2} shows an example how these components work together to make ABCO effective. As shown, removing one of the component would lead to over-prediction and decrease in performance of the model. Modeling stochastic volatility within ABCO increases its effectiveness for heterogeneous series while maintaining the same level of performance for series with constant variance. Hence, we recommend including both outlier and heteroskedastic noise when deploying ABCO, in general.

\subsection{Linear Trends / Outlier Scoring} \label{subsec_sim_lin}
Results from Section \ref{subsec_sim_sv} and Section \ref{subsec_sim_out} have shown ABCO to perform comparably well in comparison to competing methods in settings of changes in mean with outliers or stochastic volatility. One additionally advantage of ABCO is the flexible framework which allow it to detect higher-order changes and flag outliers. For this set of simulations, we will showcase both of these abilities of ABCO by simulating data with changes in linear trend and significant outliers.

Series are generated with a length of 300 with a changepoint randomly selected in the middle 20\% of the data. The linear `meet-up' model is utilized to generate the data, meaning the segments are set to be continuous (end of segment 1 becomes the start of segment 2). The start and end of segment 1 as well as end of segment two are uniformly randomly generated between -100 and 100. The segments are set to have a minimum slope difference of 1.5 to ensure separability. Three types of outlier settings are generated: small (5-10 std.\ dev.\ away from the true mean), large (25-30 std.\ dev.\ away from the true mean) and mixed (5-30 std.\ dev.\ away from the true mean). We generate N = 100 series for each outlier setting, with 5-10 outliers randomly generated for each segment. Since none of the other competing methods work well for linear trends, we ran ABCO in each of the three settings and reported the results terms of both changepoint detection accuracy and outlier detection accuracy. 

\begin{table*} [t!]
\caption{Outlier Extension: Linear Trend Results for ABCO}
\label{tab_lin}
\centering
\begin{tabular}{ c c c c c}
\hline \hline
  Outlier Type & Rand Avg. & Adj. Rand Avg. & True Positive Rate & False Positive Rate \\ 
 \hline
  Small & $0.95_{(0.01)}$ & $0.91_{(0.03)}$ & $0.71_{(0.04)}$ & $0.001_{(0.0001)}$ \\
 \hline
   Large & $0.95_{(0.02)}$ & $0.90_{(0.03)}$ & $0.95_{(0.02)}$ & $0.001_{(0.0005)}$ \\
 \hline
   Mixed & $0.94_{(0.02)}$ & $0.88_{(0.03)}$ & $0.92_{(0.02)}$ & $0.001_{(0.0002)}$ \\
  \hline
\end{tabular}
\begin{flushleft}
  \setlength{\baselineskip}{1.0pt}
{Result for ABCO on predicting both changepoints and outliers in linear settings. Average Rand and adjusted Rand reflect accuracy of changepoint detection, while true positive and false positive rates measures accuracy of outlier detection. The standard error for all measurement are shown in parentheses. }
\end{flushleft}
\end{table*}

\begin{figure} [t!]
  \begin{center}
  \caption{Linear Data Examples}
  \label{fig2L}
  \includegraphics[scale=.10]{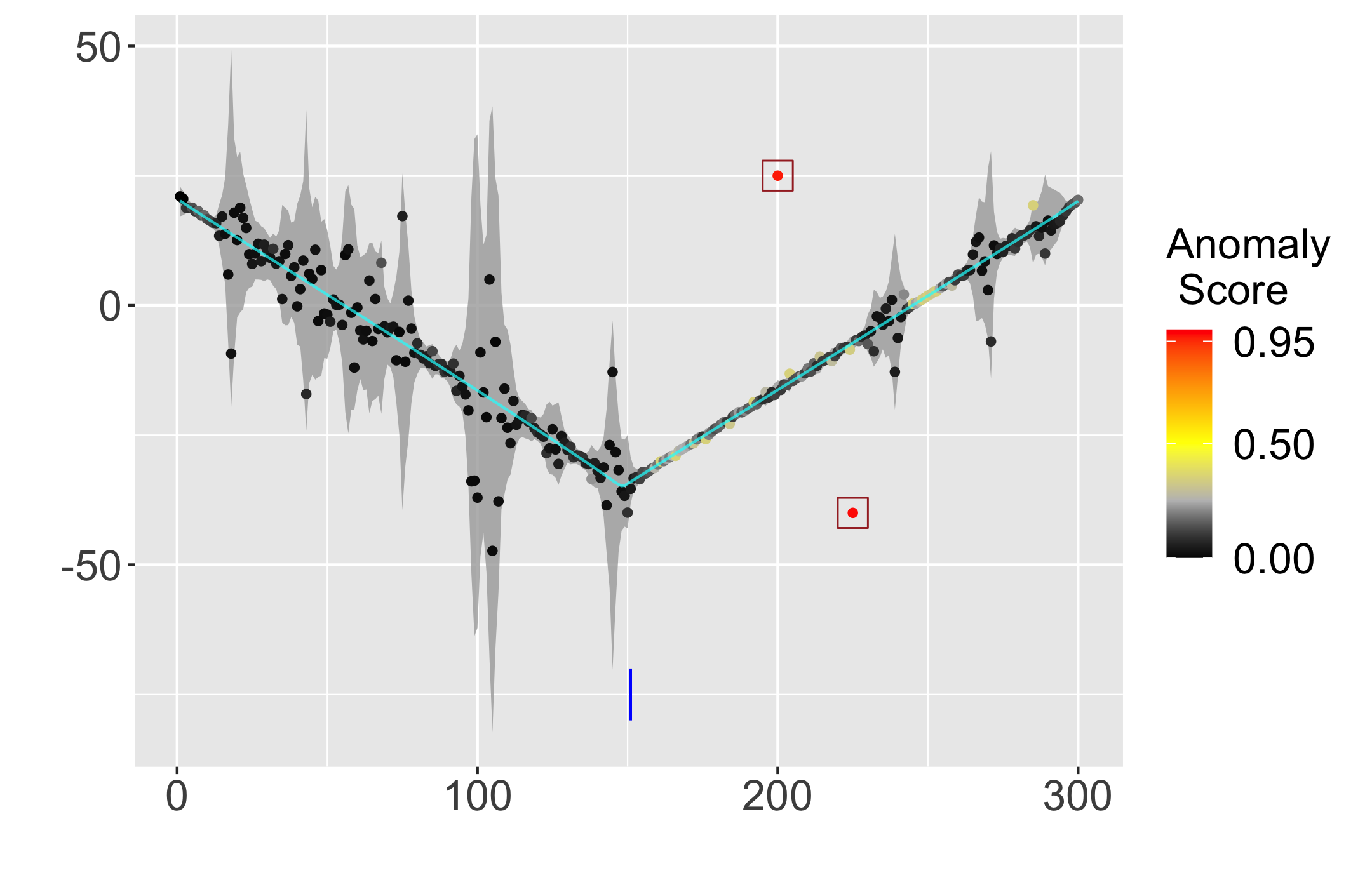} 
  \includegraphics[scale=.10]{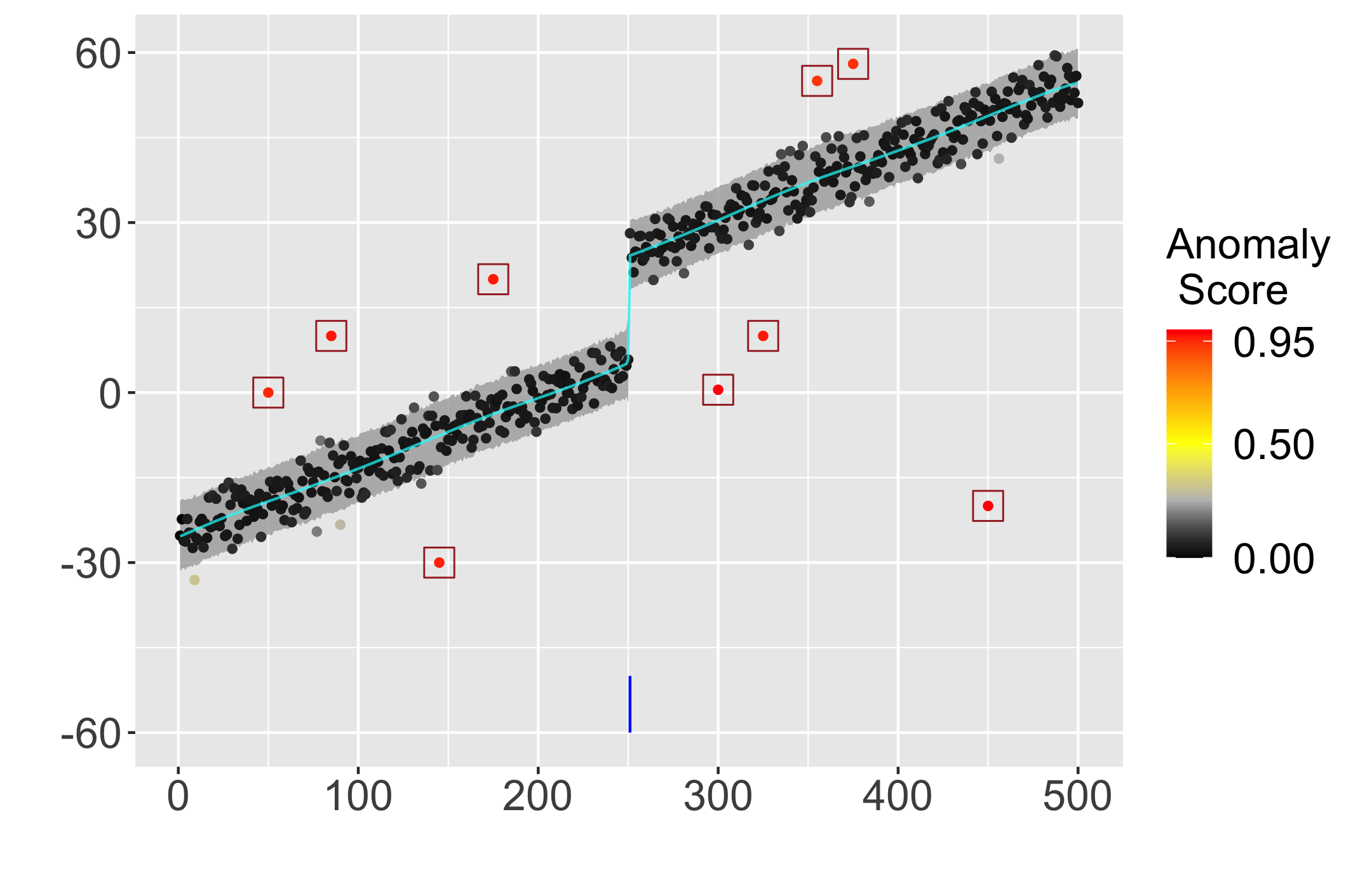} 
  \end{center}
  \begin{flushleft}
  \setlength{\baselineskip}{1.0pt}
   {Examples of plots generated using simulated series with linear trends, changepoints and outliers. Left figure shows an example of a linear meet-up model generated with stochastic volatility and 2 outliers. Right figure shows a common linear trend with a jump and several outliers. Outlier scoring is calculated by the ABCO model with true positive outliers outlined by squares. Vertical lines below the series indicate estimated changepoint locations; cyan line denote the estimated mean (trend) with 95\% credible bands for the estimated signal plus noise (excluding the estimated outlier component) in gray.}
   \end{flushleft}
\end{figure} 

For ABCO, outliers are flagged using the outlier scoring detailed in Section \ref{subsec_met_out}. We choose a cutoff threshold $0.95$ for $\{o_t\}$. 
Both true positive rate and false negative rate for outliers are reported in Table \ref{tab_lin}. As seen, ABCO is able to have a high true positive rate and a low false positive rate for all three types of outliers (an example of outlier scoring is seen in Figure \ref{fig2L}). At the same-time, ABCO is able to maintain an average adjusted Rand value of at least 0.88, signaling its effectiveness in identifying true changepoints. The detection of both changepoints and outliers simultaneously allows ABCO to provide more information for analysis. As we can see in Figure \ref{fig2L}, a series may contain a significant amount of outliers, a very difficult or impossible setting for other changepoint algorithms. The ability to incorporate both heteroskedastic noise and outlier detection makes ABCO very widely applicable. More simulation scenarios for ABCO including varying signal-to-noise ratios, changing variance and dynamic regression analysis are shown in the Online Appendix.

\section{Illustrative Applications}

This section illustrates the use of ABCO for exploring complex real data. We apply the method to the U.S. GDP and inflation. For both applications, we fit ABCO on the centered and scaled time series then transform the output back to the original scale. Further model choices are discussed within the subsequent sections.

\subsection{Gross Domestic Product Growth Rate}

For the first application we examine changepoints in gross domestic product (GDP) growth rate of the United States. Understanding changes in long-term pattern of GDP growth rate can provide important information about economic transitions \citep{antolin2017tracking}, highlight difference between the present versus the past, and infer about the effects of changes to policy \citep{levin2003inflation}. We downloaded quarterly GDP from 1947-Q1 to 2023-Q3 from the FRED database (\url{https://fred.stlouisfed.org/series/GDP}). 

\begin{figure} [ht!]
    \setlength{\tabcolsep}{2pt}
  \centering
  \caption{GDP Growth Rate Results}
  \label{fig_gdp}
   \begin{tabular}{c c}
   \includegraphics[width = 0.45\textwidth]{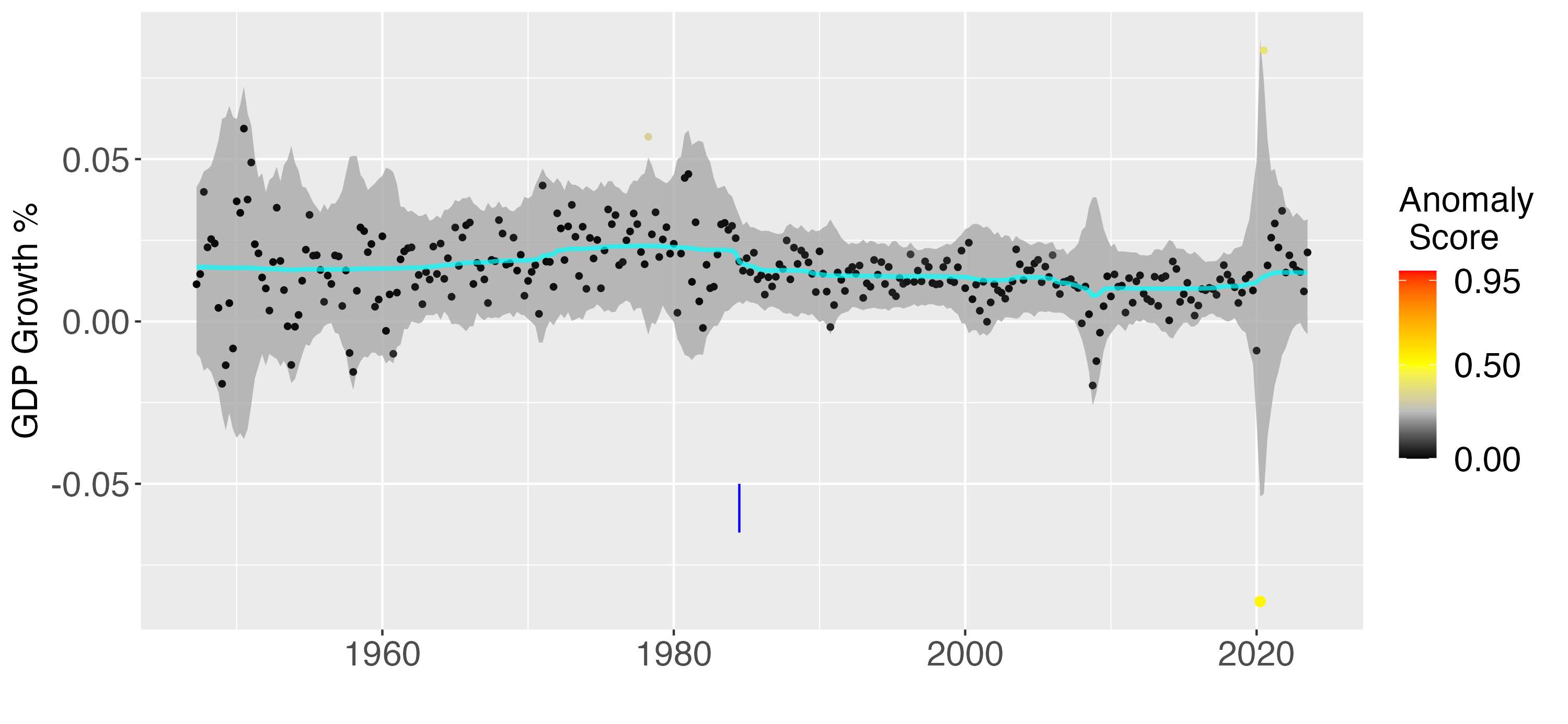} & 
   \includegraphics[width = 0.45\textwidth]{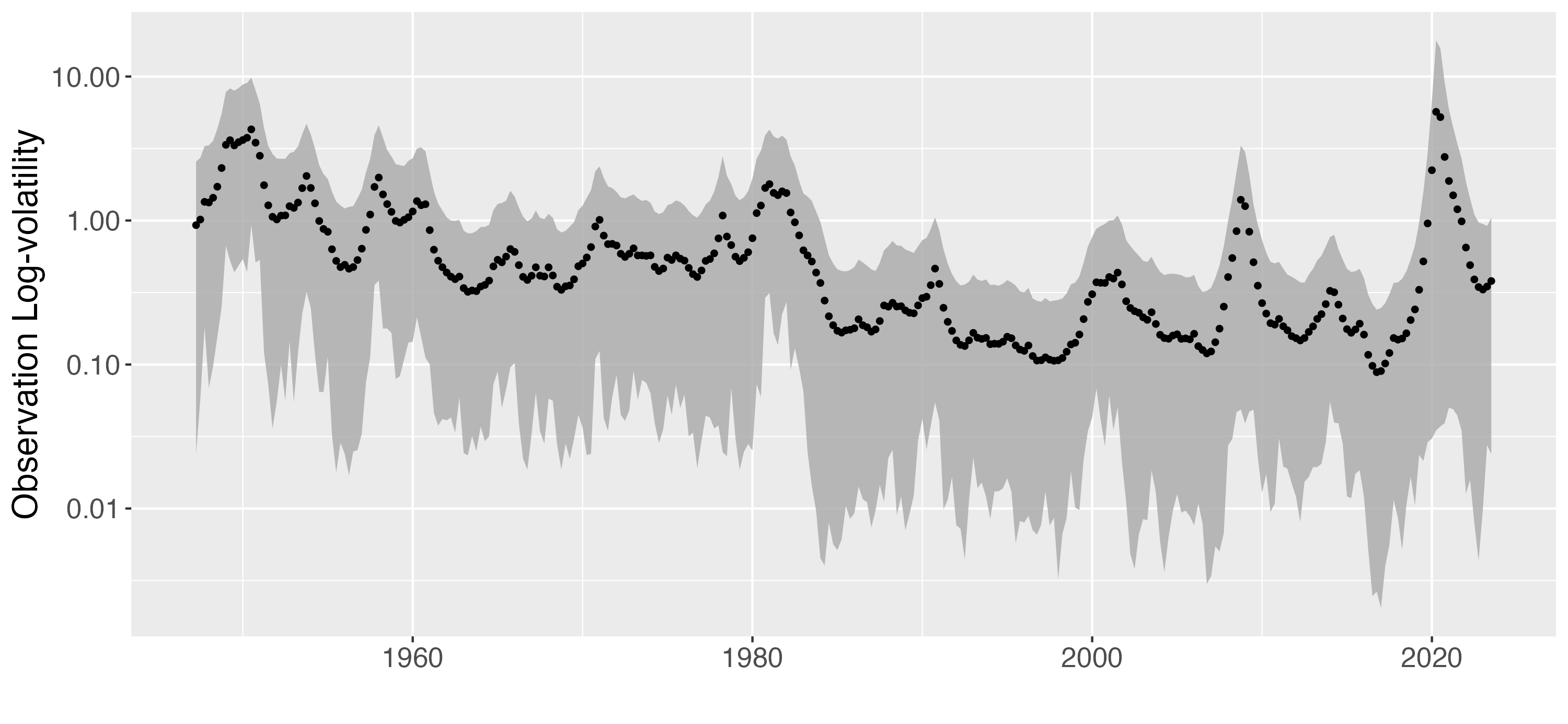} \\
   \includegraphics[width = 0.45\textwidth]{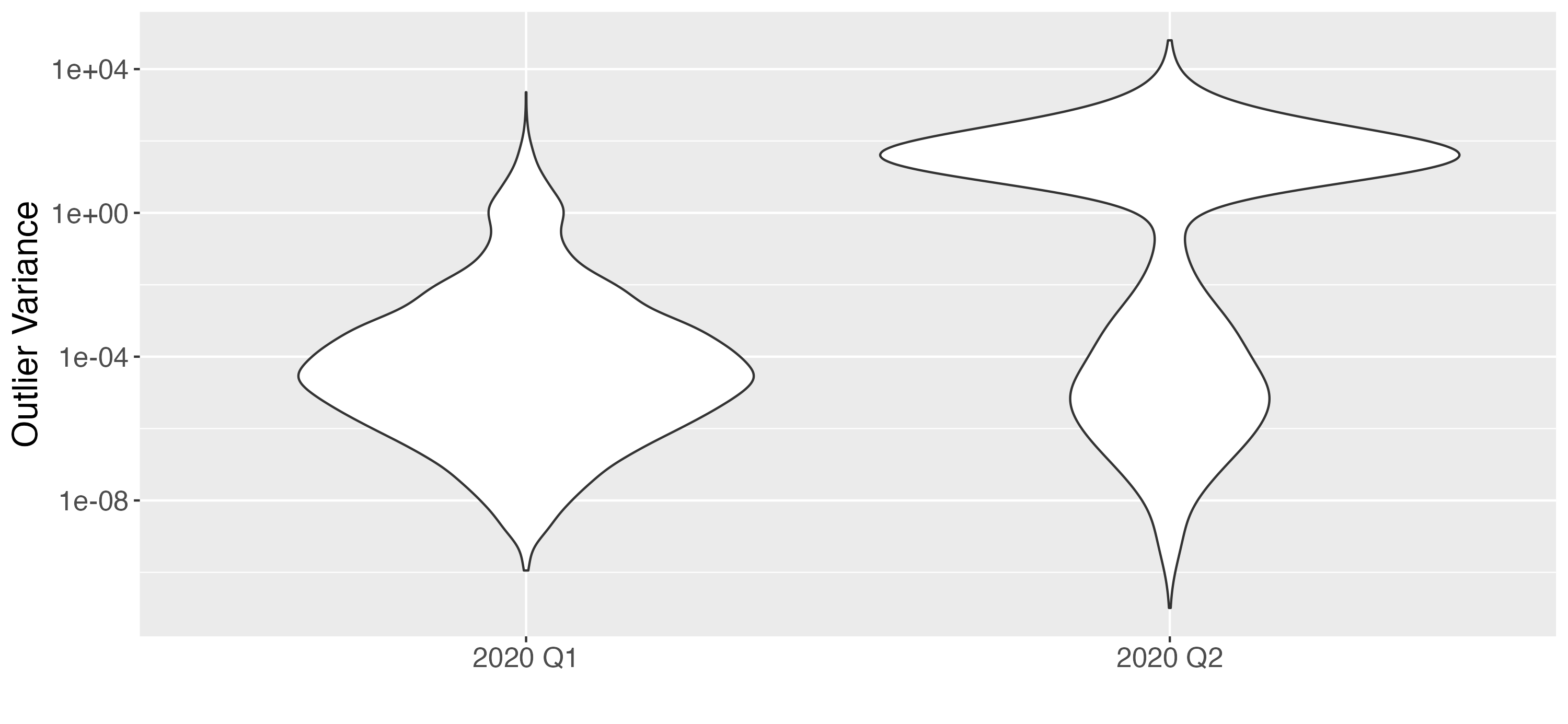} & 
   \includegraphics[width = 0.45\textwidth]{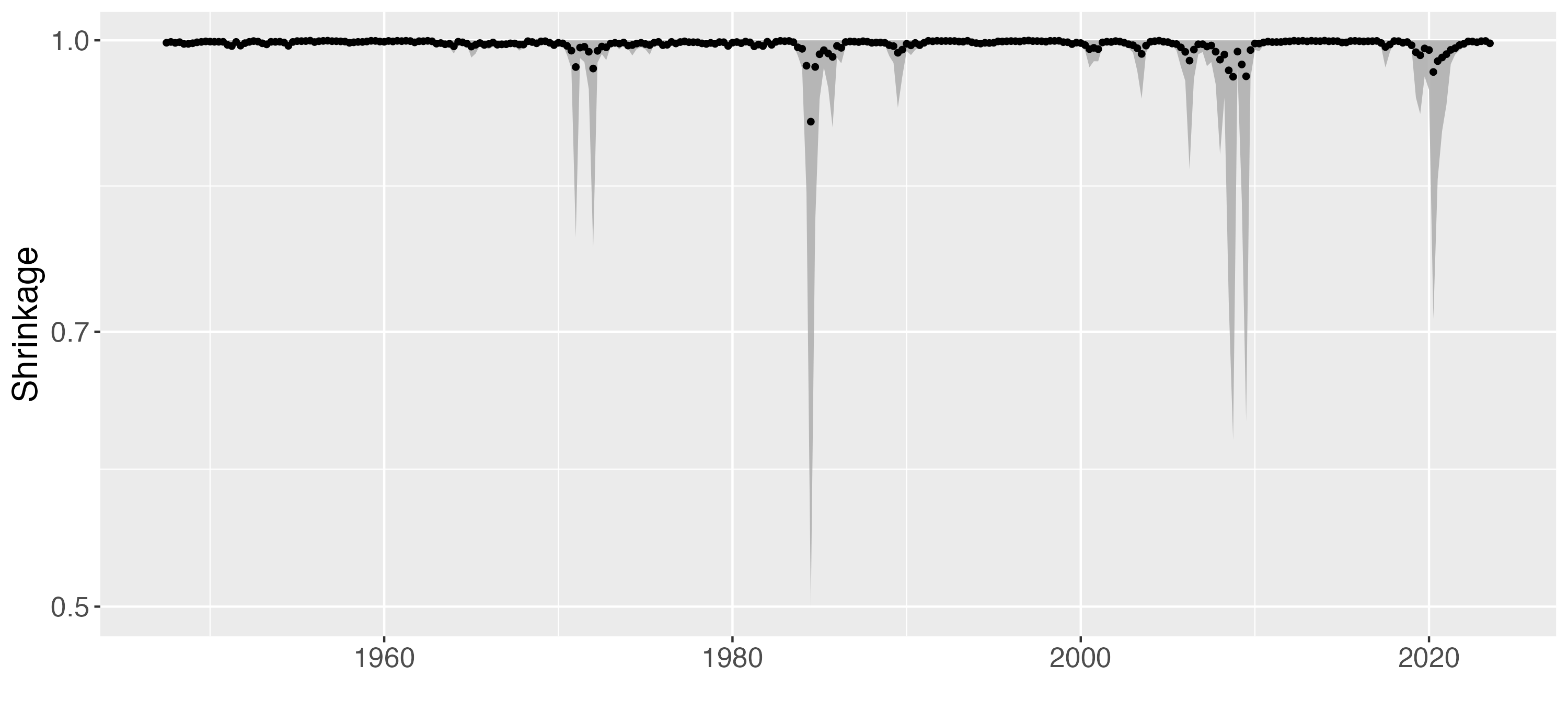} \\
  \end{tabular}
  \begin{flushleft}
  \setlength{\baselineskip}{1.0pt}
   {The top left figure depicts the fitted trend (cyan line) with 95\% prediction intervals for quarterly US GDP growth rate from 1947-2023, anomaly score, and one estimated changepoint at 1984-Q3. There is one notable outlier with an outlier score greater than 0.5 at 2020-Q2. The top right plots the estimated mean and 95\% credible intervals for the observation standard deviation on the log base 10 scale. The bottom left shows the posterior of the outlier variance for the outlying quarter, 2020-Q2, and the preceding quarter, 2020-Q1, as violin plots. The bottom right is the posterior mean and 95\% credible interval for the shrinkage proportion.}
   \end{flushleft}
\end{figure}

From inspection of the time series (Figure \ref{fig_gdp}), two features of the series make changepoint detection difficult. First, the series exhibits heteroskedastic variance with regions of high/low variability. Second, the series contain several potential outliers as a result of economic activity and social issues such as the COVID-19 pandemic. These outliers can potentially lead to over-prediction for changepoint algorithms. However, the trend is primarily characterized by changes in mean. Therefore, we fit ABCO with first differences ($D=1$) and anticipate a satisfactory fit given our observation stochastic volatility process.

Figure \ref{fig_gdp} illustrates the resulting fit of ABCO on the GDP growth series and various latent quantities are depicted. ABCO is able to account for the challenging features and predict a reasonable changepoint at 1984-Q3. In addition to the changepoint location, from the estimated observation volatility process, $\{\sigma_{\epsilon,t}\}$, in the top right panel, we can see the model estimates on average lower volatility in the mid-1980s to present. This is especially evident when looking at the lower bound of the 95\% HPD interval. This change represents a significant shift for the US economy and has been noted in previous work as an indicator of US economic stability after 1984 \citep{mcconnell1999decomposition, antolin2017tracking}. 

ABCO estimated very few outliers and only one time point had an estimated outlier score greater than 0.5, 2020-Q2. Moreover, only two additional time points had an outlier score greater than 0.3: 2020-Q3 and 1978-Q2, in order of magnitude. The bottom left panel of Figure \ref{fig_gdp} illustrates how the horseshoe+ prior on the additive outliers shapes the posterior of the outlier variance, $\lambda_{\zeta, t}^2$. The continuous mixture prior induces multimodal posteriors where one mode corresponds to the evidence of an outlier and the other corresponds to no evidence of an outlier. The sizes of the mode correspond to the strength of the evidence in the data. Overall, ABCO suggests the COVID-19 pandemic was a unique period with evidence of at least one outlier and higher volatility. However, from inspection of the last panel in the bottom left, the shrinkage profile around the time of the pandemic has minimal evidence of a change in the mean at least when compared to other periods with lower shrinkage estimates during the 1980s (our changepoint) and late 2000s. Ultimately, all of these quantities are correlated and we interpret in context given evidence of outliers is in align with previous work \citep{lenza2022how}.

\subsection{CPI Inflation} \label{sec_inf}

\begin{figure} [ht!]
    \setlength{\tabcolsep}{2pt}
  \centering
  \caption{}
  \label{fig_cpi}
   \begin{tabular}{c c}
   \includegraphics[width = 0.45\textwidth]{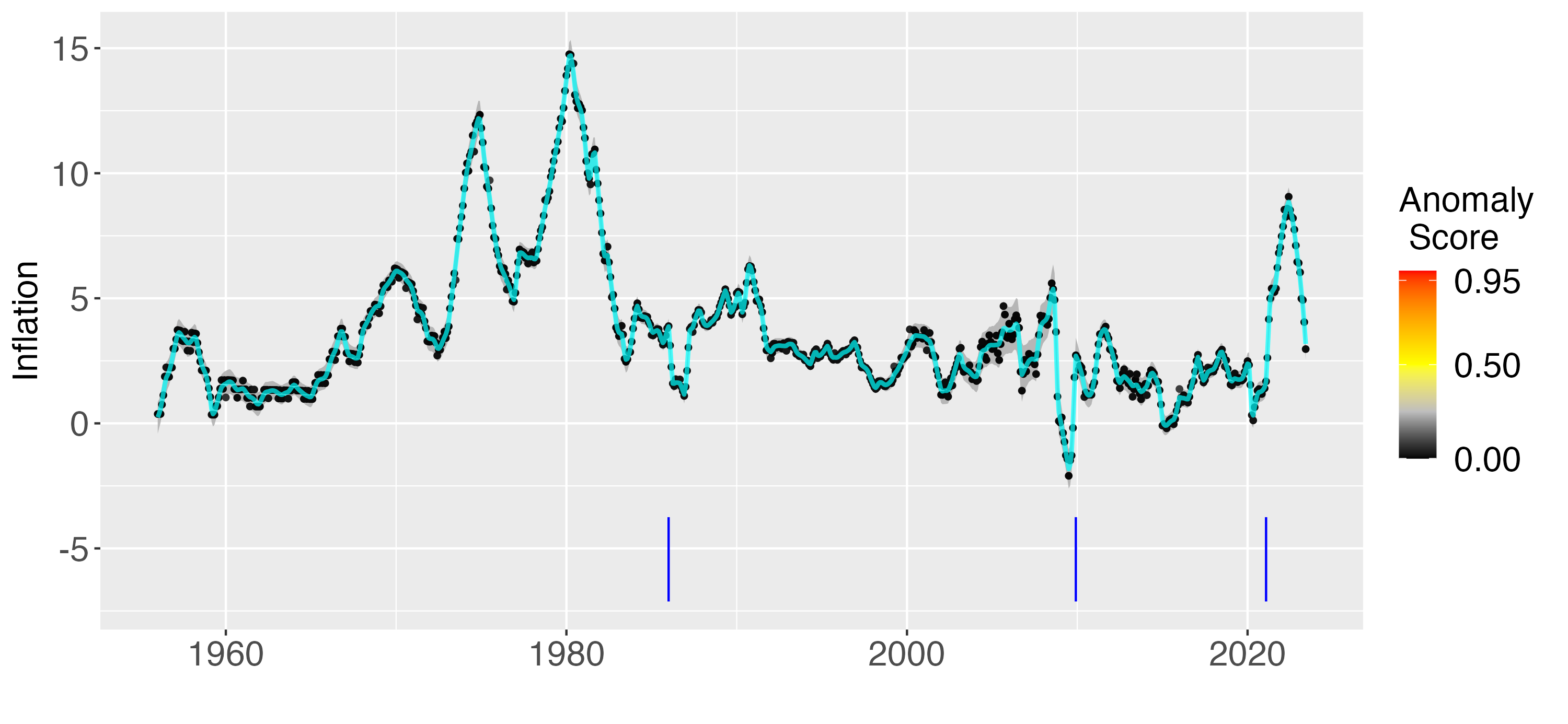} & 
   \includegraphics[width = 0.45\textwidth]{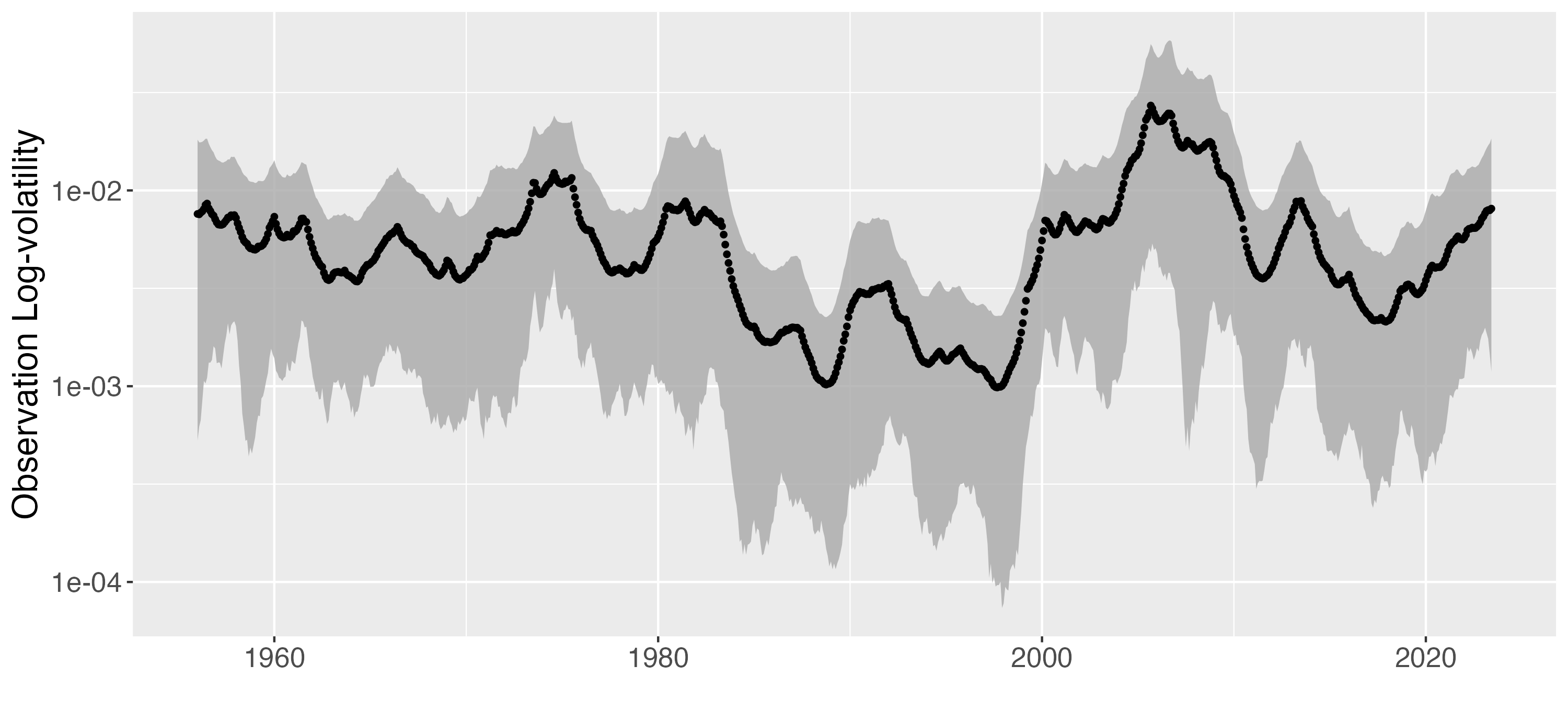} \\
   \includegraphics[width = 0.45\textwidth]{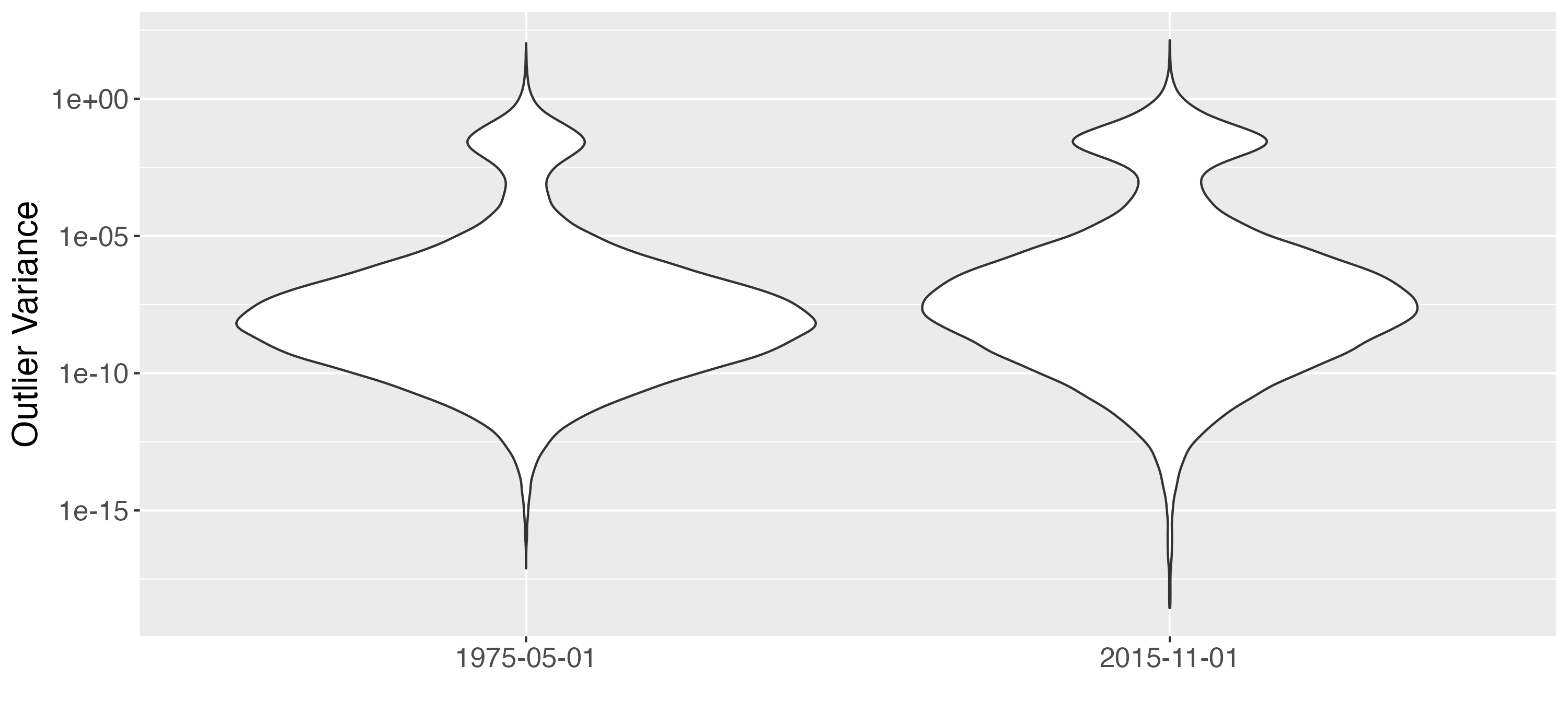} & 
   \includegraphics[width = 0.45\textwidth]{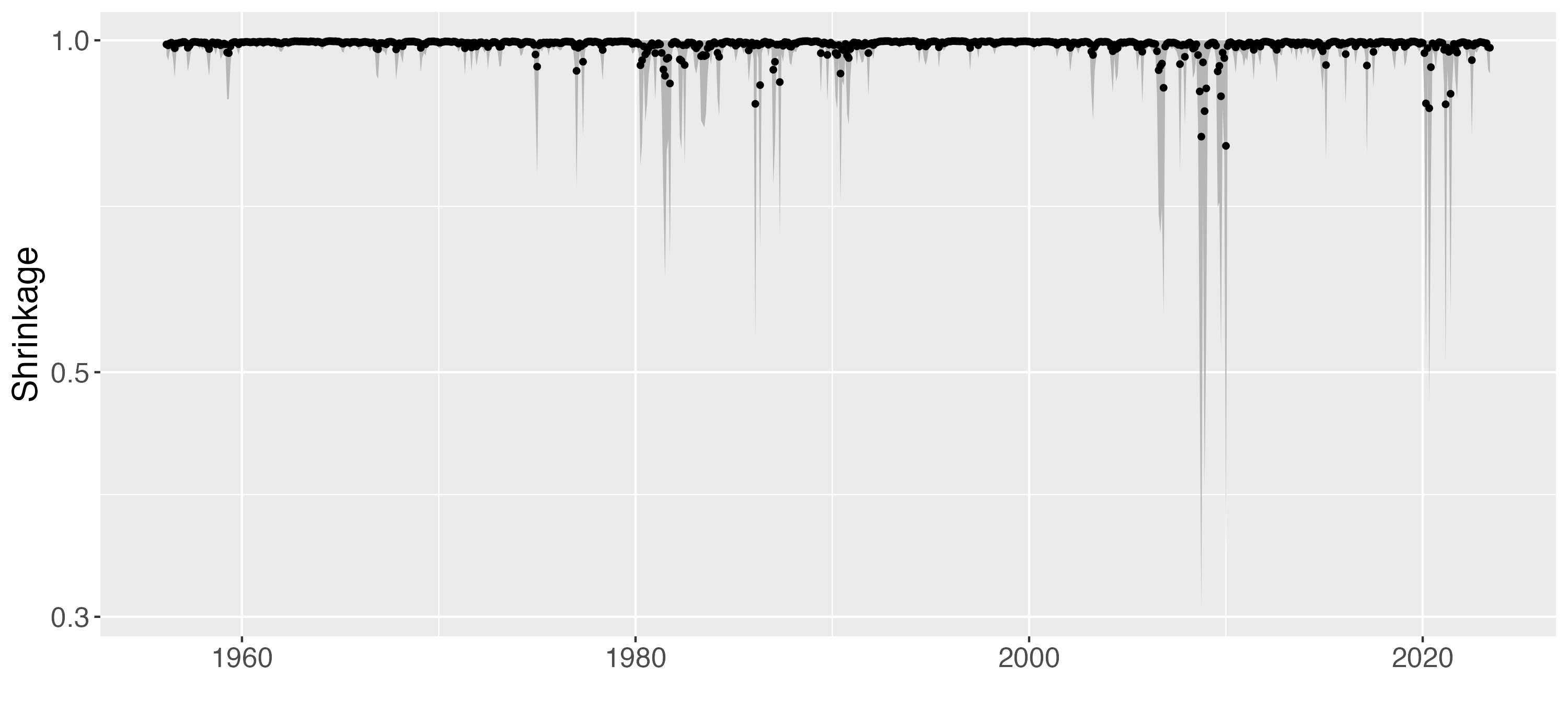} \\
  \end{tabular}
  \begin{flushleft}
  \setlength{\baselineskip}{1.0pt}
   {The top left figure depicts the fitted trend (cyan line) with 95\% prediction intervals for monthly US inflation from Jan-1956 to June-2023, anomaly score, and three estimated changepoints at Jan-1986, Dec-2009, and Feb-2021. There are no notable outliers estimated. The top right plots the estimated mean and 95\% credible intervals for the observation standard deviation on the log base 10 scale. The bottom left shows the posterior of the outlier variance for two time points, May-1975 and Nov-2015, as violin plots. The bottom right is the posterior mean and 95\% credible interval for the shrinkage proportion.}
   \end{flushleft}
\end{figure}

For the next application, we examine ABCO's capability to detect changes in linear trend by looking at the United States monthly inflation based on the CPI from Jan-1956 to June-2023 \citep{cpi}. Several previous works have explored structural break estimation for US inflation \citep{christiano2003inflation, levin2003inflation, giordani2008efficient}. As compared to the GDP, there is less heteroskedastic observation noise in the CPI inflation, but the trend is more dynamic (Figure \ref{fig_cpi}). We use the inflation data to illustrate the use of ABCO fitted on second differences ($D=2$) which includes detection of changes in the mean and changes in the rate of change in the mean.

In Figure \ref{fig_cpi}, the fitted trend is plotted with three labeled changepoints at Jan-1986, Dec-2009, and Feb-2021 that correspond to the time points with the first, second, and third highest posterior proportions of exceeding the threshold. The four segments appear to be distinguishable by a stark change in their evolution from quite dynamic to more constant or vice versa. Additionally, the top left panel of the fitted observation standard deviation suggests an increase in volatility after the 2000s, but overall volatility is quite low.

There was not strong evidence of outliers in the CPI inflation series and we have plotted the posterior densities of the two time points with the largest outlier scores. Again, the posteriors are bimodal. Lastly, notice in the posterior plot of the shrinkage proportion that the second changepoint appears to coincide with a break in the persistence of small shrinkage which reinforces our discussion from Section \ref{trend} that the TSV(1) allows for instantaneous breaks in the short-term autocorrelation of persistence.

\section{Conclusion}

We have proposed an adaptive Bayesian changepoint model with the ability to detect multiple changepoints within a time series. The model separates the series at each time-step into a trend component, an outlier component and a noise component. For trend estimation with breaks, a horseshoe-like shrinkage is placed on the $Dth$ difference of the trend component through a threshold stochastic volatility model with $Z$-distributed innovations. This shrinkage ensures most increments are relatively small or even nearly zero while allowing some isolated changes to be significantly large. The threshold variable is used to classify large changes as changepoints. For the outlier component, a horseshoe-plus prior is utilized to model extreme values in the data. For the noise component, a stochastic volatility model is specified to adapt both change and outlier detection to regions of both high and low volatility, whether stochastic or not. Together, all three components allow great flexible adaptive trend modeling, changepoint detection, and outlier scoring.  

Through simulation experiments and illustrative applications we have highlighted the unique strengths of ABCO and shown settings where it outperforms competing methods, i.e.\ changepoint series with significant outliers and/or non-constant variance. 
With a Bayesian framework, ABCO provides a reliable means for simultaneously estimating changepoints and scoring anomalies, and it can be further extended, as we demonstrated dynamic regression specifications. Additionally, the hierarchical model allows for changing specifications of different processes such as the prior without having to rederive all sampling steps. Specifically, as research on Bayesian shrinkage priors continues, one may be interested in considering alternatives to our choices. 

The labeling of time points as changepoints and outliers requires \textit{post hoc} thresholding of the latent scores, respectively. We recommend users to look at the posteriors for the outlier scores and changepoint proportions in context as we have illustrated for US GDP growth and inflation. Specifically, the posterior quantities provide an ordering such that the largest value is the most likely to be labeled. Currently, we view ABCO as a decomposition framework that significantly reduces the dimension of the unsupervised labeling of time series features to a few latent quantities with meaningful uncertainty quantification.


Further directions include analysis of multivariate series and incorporating covariates into changepoint and outlier detection equations. The challenges to these future directions include scalability and identifiability.

\bigskip
\begin{center}
{\large\bf SUPPLEMENTARY MATERIAL}
\end{center}

\begin{description}

\item[Online Appendix:] Appendix contains details about performance and more simulations comparing effectiveness of ABCO against other changepoint algorithms.
\item[Code:] Zip file containing the code for running ABCO. The zip file includes all sampling algorithms for ABCO. It also includes code for running all simulations and real world examples seen in the paper. 
\end{description}

\bibliographystyle{agsm}
\bibliography{paper}

\end{document}


\def\spacingset#1{\renewcommand{\baselinestretch}%
{#1}\small\normalsize} \spacingset{1}

\maketitle

\newpage
\spacingset{1.5}

\section{Further Simulation Details and Settings}

The Rand index measures similarity between two different segmentations of a series. Let $X$ denote a segmentation given by a  model and let $Y$ denote the true segmentation of the series. All pairs of points can then divided into four groups: pairs that are placed in the same segment in $X$ and in same segment in $Y$ (A); pairs that are placed in same segment in $X$ and in different segments in $Y$ (B); pairs that are placed in different segments in $X$ and in same segment in $Y$ (C); pairs that are placed in different segments in $X$ and in different segments in $Y$ (D). The Rand index is calculated as $\frac{A+D}{A+B+C+D}$. The Adjusted Rand index corrects for the random chance that pairs of points will be placed together. We utilize the Rand index and adjusted Rand index as they give a simple but holistic measure of similarity between the estimated segmentation and the true segmentation. 

\subsection{Performance}

\begin{figure}[ht!]
    \centering
    \includegraphics[width = \textwidth]{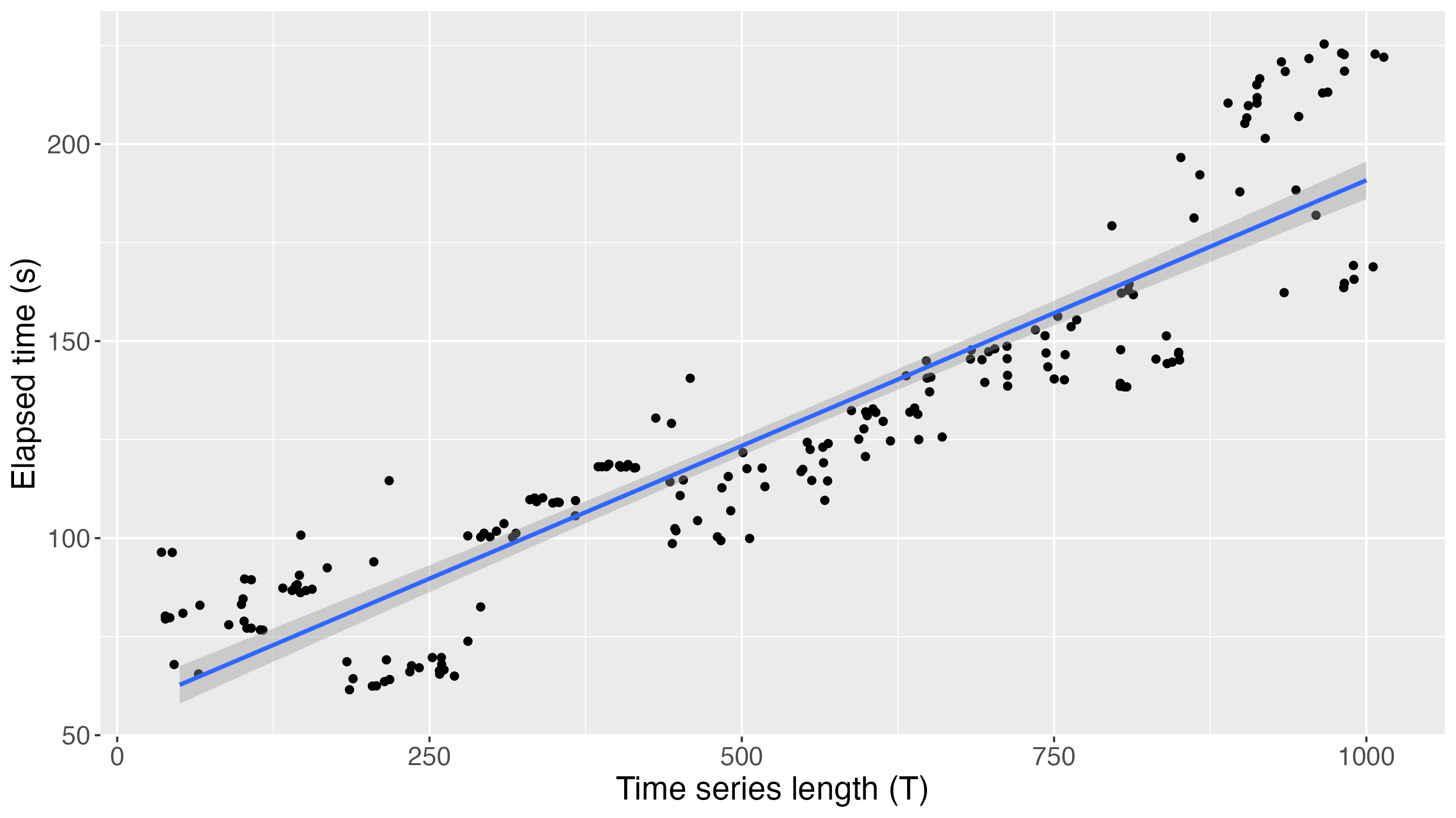}
    \caption{Scatterplot of elapsed time for 10000 MCMC samples (5000 burnin and 5000 post) to a dataset of varying time series length (T). For each time series length, the sampling was repeated 10 times. A simple linear regression fit with standard error is overlaid. }
    \label{fig:time-trials}
\end{figure}

\begin{figure}[ht!]
    \centering
    \includegraphics[width = \textwidth]{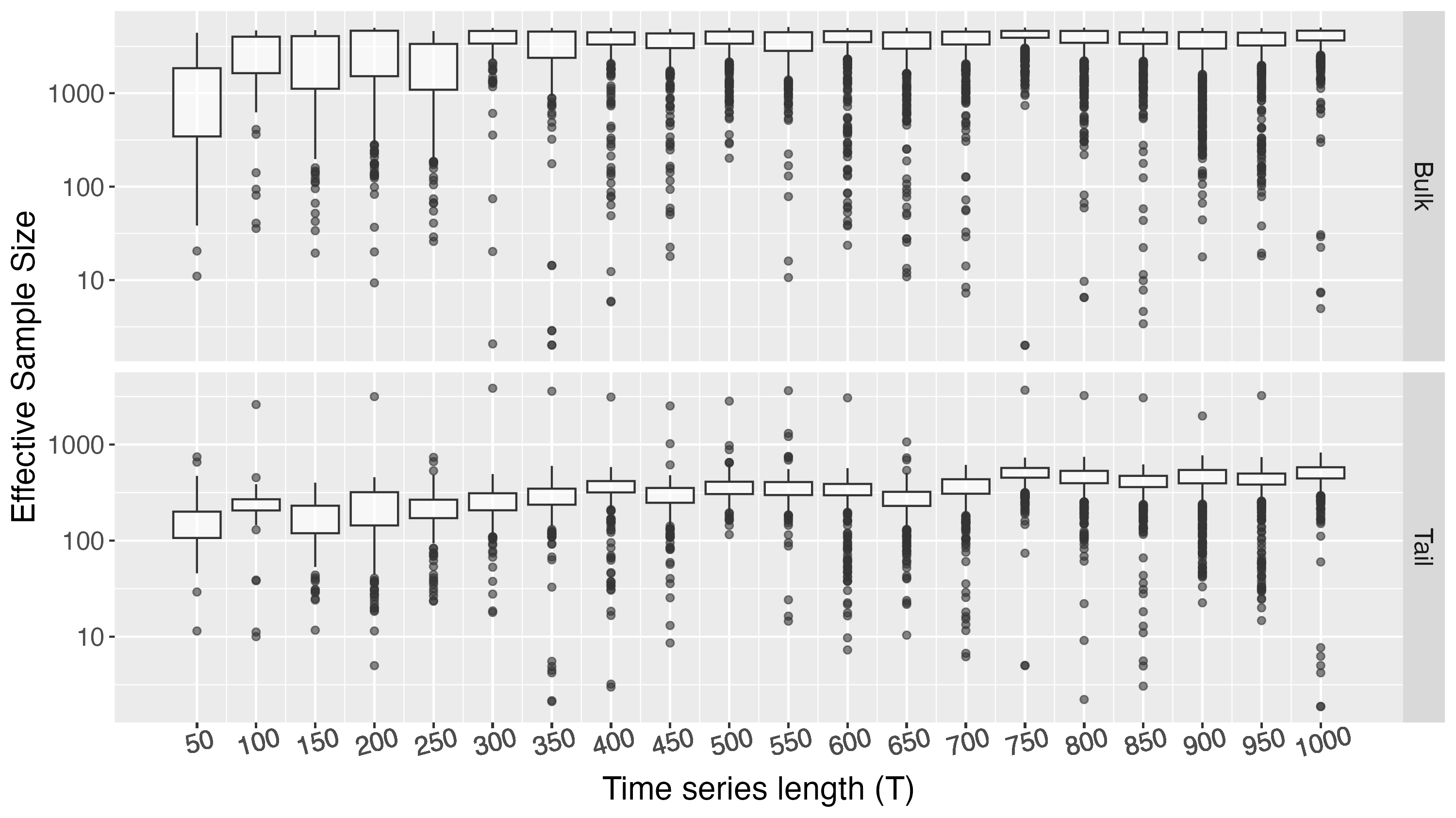}
    \caption{Boxplots of bulk and tail effective sample sizes across 10 chains of 10000 MCMC samples (5000 burnin and 5000 post) to a dataset of varying time series length (T).}
    \label{fig:ess}
\end{figure}

In order to assess the scalability and convergence of the proposed MCMC algorithm, we repeated the model fitting ten times to the simulation setting described in Section 4.1 of the main text. For each model fit, the MCMC algorithm was run for 10000 iterations (5000 burnin and 5000 post).The empirical results suggest the algorithm scales linear with time (Figure \ref{fig:time-trials}) which is consistent with the theory presented in \citet{kowal_2018}. For convergence diagnostics, we considered the $\hat{R}$ (ratio of between and within chain variance) and bulk and tail effective sample sizes. Figure \ref{fig:ess} shows that majority of the chains have an effect sample size of at least 100 for the posterior bulk and tail recommended for posterior mean and quantile summaries. Most rules of thumb suggest a limit of $\hat{R}$ between 1.05 and 1.1. Based on Table \ref{tab:rhat_sim}, majority of the settings have maximum $\hat{R}$ values within that range. For those settings that do not meet the rules of thumb, it suggests running the MCMC for longer iterations. However, $\hat{R}$ does have known limitations for multimodal posteriors which are common with shrinkage priors. All diagnostic criteria were calculated using the \textit{rstan} R package \citep{rstan}. 

\begin{table}[!h]
\centering
     \caption{Range of $\hat{R}$ as Time Series Length Varies}
    \begin{tabular}{rrr}
    \hline
    \hline
    T & $\hat{R}$ Min & $\hat{R}$ Max\\
    \hline
    50 & 1.00 & 1.11\\
    \hline
    100 & 1.00 & 1.06\\
    \hline
    150 & 1.00 & 1.10\\
    \hline
    200 & 1.00 & 1.11\\
    \hline
    250 & 1.01 & 1.06\\
    \hline
    300 & 1.00 & 1.41\\
    \hline
    350 & 1.00 & 1.43\\
    \hline
    400 & 1.00 & 1.12\\
    \hline
    450 & 1.00 & 1.08\\
    \hline
    500 & 1.00 & 1.03\\
    \hline
    550 & 1.00 & 1.06\\
    \hline
    600 & 1.00 & 1.06\\
    \hline
    650 & 1.00 & 1.13\\
    \hline
    700 & 1.00 & 1.11\\
    \hline
    750 & 1.00 & 1.43\\
    \hline
    800 & 1.00 & 1.11\\
    \hline
    850 & 1.00 & 1.25\\
    \hline
    900 & 1.00 & 1.06\\
    \hline
    950 & 1.00 & 1.07\\
    \hline
    1000 & 1.00 & 1.13\\
    \hline
    \end{tabular}
    \begin{flushleft}
    \setlength{\baselineskip}{1.0pt}
    {Range of the $\hat{R}$ for 10 chains of 5000 samples fit to time series of various length (T) for the $\{\omega_t\}$ and $r$ parameters.}
    \end{flushleft}
    \label{tab:rhat_sim}
\end{table}

\clearpage

\subsection{Multiple Changes in Mean with Heteroskedastic Noise}
This subsection shows additional results using similar setting as Section 3.1. We generate $N=100$ simulated series of length $T=1000$. For each series, the number of changepoints are generated uniformly at random from the set $\{2, 3, 4\}$. Stochastic volatility are added to the mean signals, with log volatility specified as \begin{equation}
\log(\sigma_{\epsilon, t}^2) = \phi_\epsilon \log(\sigma_{\epsilon, t-1}^2) + \nu_{\epsilon, t},  \; \; \; \; \; \; \; \; \; \; \; \; \nu_{\epsilon, t} \sim N(0, \sigma_\nu^2),
\end{equation}
with $\phi_\epsilon = 0.9$. We change the variable $\sigma_\nu^2$ in order to illustrate the performance of the algorithms in varying signal-to-noise ratios. We add two additional settings of $\sigma_\nu^2 = \{1.25, 1.5\}$. The results are shown in Table \ref{tab_sv_add1} and \ref{tab_sv_add2}.

\begin{table*}[ht!]
\caption{Mean Changes with Stochastic Volatility, $\sigma_\nu^2 = 1.25$}
\label{tab_sv_add1}
\centering
\begin{tabular}{ c c c c c c c}
\hline \hline
  Algorithms &  Rand Avg. & Adj. Rand Avg. & F1-Score & Avg. Dist. to True \\ 
 \hline
 ABCO & $0.94_{(0.01)}$ & $0.89_{(0.02)}$ & $\textbf{0.85}$ & $\textbf{0.86}_{(0.22)}$ \\
 \hline
 Horseshoe & $0.63_{(0.02)}$ & $0.25_{(0.03)}$ & $0.08$ & $127.62_{(14.81)}$ \\
 \hline
 R-FPOP & $0.93_{(0.01)}$ & $0.86_{(0.02)}$ & $0.73$ & $9.83_{(2.56)}$ \\
 \hline 
 E.Divisive & $0.82_{(0.01)}$ & $0.63_{(0.02)}$ & $0.51$ & $48.83_{(4.24)}$ \\
 \hline
 H-SMUCE & $\textbf{0.95}_{(0.01)}$ & $\textbf{0.90}_{(0.02)}$ & $0.71$ & $13.12_{(3.46)}$ \\
 \hline
\end{tabular}
\begin{flushleft}
  \setlength{\baselineskip}{1.0pt}
{Results for additional simulations for mean changes with stochastic volatility with setting $\sigma_\nu^2 = 1.25$.}
\end{flushleft}
\end{table*}

\begin{table*}[ht!]
\caption{Mean Changes with Stochastic Volatility, $\sigma_\nu^2 = 1.5$}
\label{tab_sv_add2}
\centering
\begin{tabular}{ c c c c c c c}
\hline \hline
  Algorithms & Rand. Avg. & Adj. Rand Avg. & F1-Score & Avg. Dist. to True \\ 
 \hline
 ABCO  & $0.94_{(0.01)}$ & $0.88_{(0.02)}$ & $\textbf{0.81}$ & $\textbf{1.05}_{(0.22)}$ \\
 \hline
 Horseshoe & $0.56_{(0.02)}$ & $0.18_{(0.03)}$ & $0.05$ & $161.61_{(18.62)}$ \\
 \hline
 R-FPOP  & $0.93_{(0.01)}$ & $0.85_{(0.02)}$ & $0.70$ & $10.89_{(2.64)}$ \\
 \hline 
 E.Divisive &  $0.82_{(0.01)}$ & $0.63_{(0.02)}$ & $0.50$ & $68.83_{(4.76)}$ \\
 \hline
 H-SMUCE & $\textbf{0.95}_{(0.01)}$ & $\textbf{0.90}_{(0.02)}$ & $0.64$ & $14.22_{(2.19)}$ \\
 \hline
\end{tabular}
\begin{flushleft}
  \setlength{\baselineskip}{1.0pt}
{Results for additional simulations for mean changes with stochastic volatility with setting $\sigma_\nu^2 = 1.5$.}
\end{flushleft}
\end{table*}

As seen in the results, the Rand average and adjusted Rand average did not change much for ABCO, R-FPOP and H-SMUCE as signal-to-noise ratios increase. While H-SMUCE slightly out-performs ABCO in terms of average adjusted Rand, ABCO's result remains within one standard error. However, comparing the F1-score and average distance to true, we can see the gap between ABCO and H-SMUCE widens as signal-to-noise ratio increases. This shows robustness of ABCO to noisy dataset. ABCO remains the most effective at identifying the correct changepoint locations and maintains the best balance of precision vs recall in all signal-to-noise settings.

\subsection{Robustness to Changes in Variance}
For this set of simulations, we consider the effectiveness of ABCO in estimating changepoints in mean in presence of changing variance. We generate $N = 100$ series of length $T = 300$. Each series is composed of two segments, with a changepoint in the middle. The first segment has a mean of 0 and Gaussian noise with variance 1. The second segment has a mean of 5 and Gaussian noise with variance $\sigma_v^2$. The variance for the second term is selected from $\sigma_v^2 = \{2, 5, 10\}$. By having the variance also change between the segments, this will illustrate the robustness of ABCO. 
\begin{table}[ht!]
\caption{Changepoints in Mean with Changing Variance}
\label{tab_ch_var}
\centering
\begin{tabular}{ c c c c c c}
\hline \hline
 Method & $\sigma_v^2$ & Rand. Avg. & Adj. Rand Avg. & F1-Score & Avg. Dist. to True \\
 \hline
 ABCO & 2 & $0.99_{(0.01)}$ & $0.99_{(0.01)}$ & 0.89 & $0.03_{(0.01)}$  \\ 
  & 5 &  $0.97_{(0.01)}$ & $0.94_{(0.02)}$ & 0.83 & $0.48_{(0.12)}$  \\
 & 10 &   $0.94_{(0.02)}$ & $0.88_{(0.04)}$ & 0.76 & $0.82_{(0.17)}$ \\
 \hline
\end{tabular}
\begin{flushleft}
  \setlength{\baselineskip}{1.0pt}
{Results of ABCO in presence of changing variance. The column $\sigma_v^2$ indicates the Gaussian noise for the second segment. We show three different magnitude of changes with $\sigma_v^2 = \{2, 5, 10\}$.}
\end{flushleft}
\end{table} 

The results are shown in Table \ref{tab_ch_var}. As seen, ABCO is able to perform well in all 3 settings of changing variance. In the setting of $\sigma_v^2 = 2$, ABCO maintains an adjusted Rand average of $0.99$ and a F1-score of 0.89. In the setting of $\sigma_v^2 = 10$, ABCO maintains an adjusted Rand average of $0.88$ and a F1-score of 0.76. This shows that ABCO is robust to changes in variance and illustrates the effectiveness of the Bayesian framework. In all 3 settings, the average distance from predicted changepoint to true changepoint remains less than 1. This shows ABCO's accuracy in its predictions. With a heteroskedastic noise component and an outlier component, ABCO maintains local adaptivity and can identify changepoints in mean with changing variance. 

\subsection{Dynamic Regression Simulation Scenarios}
For dynamic regression simulations, the goal is to test ABCO-X's ability to detect changes in the coefficients of the predictors. We simulated 100 series of lengths 100, 200, 400 with Gaussian noise, with variance 1. For each series, we generate three predictors ($\pmb{x_t} = [x_{1,t}, x_{2,t}, x_{3,t}]$): the first predictor ($\{x_{1,t}\}$) is a vector of 1s; the second predictor ($\{x_{2,t}\}$) and the third predictor ($\{x_{3,t}\}$) are iid standard Gaussian. For the time-varying coefficient matrix $\{\pmb{\beta_t}\}$, the coefficients for the first predictor is piecewise constant, with jumps (two changepoints) in the middle 50th percentile of the data, while the coefficients for the second and third predictors are random walks with iid increments. The response variable is given as
$$
y_t = \pmb{x_t'}\pmb{\beta}_t + \epsilon_t, \; \; \; \; \epsilon_t \sim N(0,1).
$$
In this dynamic regression scenario, we only ran ABCO-X on the data as most other changepoint algorithms are not adaptable to the regression setting. For each simulation, we calculated the average adjusted Rand index for the first predictor as well as the average number of changepoints predicted for each predictor. The results are shown in Table \ref{tab3}.
\begin{table}[t!]
\caption{Multivariate Simulation with Three Predictors}
\label{tab3}
\centering
\resizebox{\columnwidth}{!}{
\begin{tabular}{ c c c c c c c}
\hline \hline
 Method & Length (T) & Adj. Rand Avg. & Avg. CP Pred1 & Avg. CP Pred2 & Avg. CP Pred3\\ 
 \hline
 ABCO-X & 100 & 0.9928 & 2.34 & 0 & 0\\ 
  & 200 & 0.9856 & 2.44 & 0 & 0 \\
 & 400 & 0.9665 & 2.67 & 0 & 0 \\
 \hline
\end{tabular}
}
\begin{flushleft}
  \setlength{\baselineskip}{1.0pt}
{Avg.\ CP Pred1 measures on average how many changepoints are estimated in the coefficient series for predictor 1, and similarly for Pred2 and Pred3. Adjusted Rand Avg.\ is calculated only for the coefficient of the first predictor.}
\end{flushleft}
\end{table} 

As shown in Table \ref{tab3}, ABCO-X is very accurate for multivariate examples. The algorithm correctly detects changepoints only in the first coefficient series (predictor 1) while detecting no changepoints in the other coefficient series. The ability to find component specific changes with respect to $\{\beta_t\}$ can be very useful for identifying which predictors most strongly associate with changes in the response variable over time. Across various series lengths, ABCO-X consistently achieves an average adjusted Rand value of above 0.95, showing it can consistently detect the true location of changepoints even across smaller segments. 

\begingroup
\setstretch{1.4}
\setlength{\bibsep}{4pt}
\bibliographystyle{agsm}
\bibliography{paper}
\endgroup